\documentclass{elsart}               % The use of LaTeX2e is preferred.

\usepackage{graphicx}
\usepackage{epsfig}  
\usepackage{amsmath}
\usepackage{indentfirst}
\usepackage{algorithm}
\usepackage{algorithmic}   
\usepackage{color}
\usepackage[dvipsnames]{xcolor}
\usepackage{amssymb} 
\usepackage{subcaption}
\usepackage{tikz}
\usepackage{soul}
\usepackage{xcolor}
\usepackage{mathtools}
\usepackage{optidef}
\usepackage{breqn}
\usepackage{cite}
\usepackage[algo2e,norelsize]{algorithm2e} 

\SetKwInput{kwAdd}{Add}
\SetKwInput{kwSet}{Set}
\SetKwInput{kwReset}{Reset}
\SetKwInput{kwUpdate}{Update}
\SetKwInput{kwClear}{Clear}
\SetKwInput{kwSave}{Save}
\SetKwInput{kwBuild}{Build}
\SetKwInput{kwIni}{Initialization}
\usetikzlibrary{matrix,chains,positioning,decorations.pathreplacing,arrows}

\newtheorem{definition}{Definition}
\newtheorem{lemma}{Lemma}
\newtheorem{remark}{Remark}

\newtheorem{assumption}{Assumption}

\newtheorem{corollary}{Corollary}%[theorem]

\newcommand{\sgn}{\text{sgn}}

\newcommand{\bG}{{\mathbf G}}

\newcommand{\bX}{{\mathbf X}}

\newcommand{\bL}{{\mathbf L}}

\newcommand{\bI}{{\mathbf I}}

\def\matt#1{\begin{bmatrix}#1\end{bmatrix}}

\begin{document}

\begin{frontmatter}

% \title{Deep Koopman Learning of Nonlinear Time-Varying Systems \thanksref{footnoteinfo}} % Title, preferably not more 
\title{Deep Koopman Learning of Nonlinear Time-Varying Systems}

% \author[1]{Wenjian Hao}\ead{hao93@purdue.edu},
% \author[2]{Bowen Huang}\ead{bowen.h@pnnl.gov},
% \author[3]{Wei Pan}\ead{wei.pan@manchester.ac.uk},
% \author[2]{Di Wu}\ead{di.wu@pnnl.gov},
% \author[1]{Shaoshuai Mou}\ead{mous@purdue.edu}

\author[1]{Wenjian Hao},
\author[2]{Bowen Huang},
\author[3]{Wei Pan},
\author[2]{Di Wu},
\author[1]{Shaoshuai Mou}

\address[1]{School of Aeronautics and Astronautics, Purdue University, West Lafayette, USA}
\address[2]{Pacific Northwest National Laboratory, Richland, USA}
\address[3]{Department of Computer Science, University of Manchester, UK}

\thanks[footnoteinfo]{Email addresses: \texttt{\{hao93, mous\}@purdue.edu}, \texttt{\{bowen.h, di.wu\}@pnnl.gov}, \texttt{wei.pan@manchester.ac.uk}.}
          
                            % keyword list or with the 
                                          % help of the Automatica 
                                          % keyword wizard

\begin{abstract}
This paper presents a data-driven approach to approximate the dynamics of a nonlinear time-varying system (NTVS) by a linear time-varying system (LTVS), which results from the Koopman operator and deep neural networks. Analysis of the approximation error between states of the NTVS and the resulting LTVS is presented. Simulations on a representative NTVS show that the proposed method achieves small approximation errors, even when the system changes rapidly. Furthermore, simulations in an example of quadcopters demonstrate the computational efficiency of the proposed approach.
\end{abstract}
\end{frontmatter}

\section{Introduction}
In recent years, data-driven methods have received a significant amount of research attention due to the increasing complexity of the autonomous systems in both dynamics \cite{mamakoukas2021derivative,mezic2015applications, proctor2018generalizing, mauroy2016linear} and mission objectives \cite{WDJSS19TRO,WS21Auto,WDSS21IJRR}. In the direction of learning system dynamics, the Koopman operator has recently been proven to be an effective method to approximate a nonlinear system by a linear time-varying system based on state-control pairs \cite{mezic2015applications, proctor2018generalizing, mauroy2016linear}. Along this direction, two popular methods dynamic mode decomposition (DMD) and extending dynamic mode decomposition (EDMD) are used to lift the state space to a higher-dimensional space, where the evolution is approximately linear \cite{korda2018linear}. However, choosing the proper observable functions for the lifting transformation is still an open question, and the potentially large lifted dimension may hinder real-time applications.

Recent work has proposed several methods for choosing proper observable functions of Koopman-based methods for time-invariant systems. Lusch et al. \cite{lusch2017data} proposed applying deep learning methods to discover the eigenfunctions of the approximated Koopman operator. Yeung et al. \cite{yeung2019learning, dk2, dk, bevanda2021koopmanizingflows} introduced deep neural networks (DNN) as observable functions of the Koopman operator, which are tuned based on collected state-control pairs by minimizing a properly defined loss function. While some work, such as \cite{zhang2019online}, has extended the DMD method to approximate nonlinear time-varying systems (NTVS) that change sufficiently slowly by linear time-varying systems (LTVS), this method is not directly applicable to approximate nonlinear systems with rapidly changing dynamics.

In this paper, we propose a deep Koopman learning method to approximate NTVS, which employs DNN as the observable function of the Koopman operator and adjusts both the DNN and the approximated dynamical system simultaneously. This is achieved by tuning the DNN parameters based on the latest state-control data pairs to track the unknown NTVS. 

Compared to existing results in \cite{zhang2019online}, the proposed method is able to approximate an NTVS which does not necessarily change slowly.  
Contributions of this paper are summarized as follows:
\begin{itemize}
\item We propose a deep Koopman representation formulation for the NTVS and provide a practical online algorithm for implementation.
\item We investigate the error bound of the system state estimation of the proposed method.
\item We perform a convergence analysis of the proposed method concerning the observable function of DNN.
\end{itemize}
% }
This paper is organized as follows. In Section \ref{pformulation}, we state the problem. Section \ref{pmethod} presents the main results. The numerical simulations are exhibited in Section \ref{results}. Finally, Section \ref{conclusion} concludes the paper.
 
\noindent{\emph{Notations}}. Let $\parallel \cdot \parallel$ denote the Euclidean norm. For a matrix $A\in\mathbb{R}^{n\times m}$, $\parallel A \parallel_F$ denotes its Frobenius norm; $A^T$ denotes its transpose; $A^\dagger$ denotes its Moore-Penrose pseudoinverse.
For positive integers $n$ and $m$, $\bI_{n}$ denotes the ${n \times n}$ identity matrix; $\mathbf{0}_n \in\mathbb{R}^n$ denotes a vector with all value $0$; $\mathbf{0}_{n\times m}$ denotes a $n\times m$ matrix with all value $0$. $\sgn{(\cdot)}$ denotes the sign function. $\lceil \cdot \rceil$ denotes the ceiling function, i.e., given real numbers $y$, integers $k$ and the set of integers $\mathbb{Z}$, $\lceil y \rceil = \min\{k\in\mathbb{Z}\ |\ k\geq y\}$.

\section{Problem Formulation}\label{pformulation}
Consider an NTVS, the dynamics of which is unknown. Let $x_t\in \mathbb{R}^n$ and $u_t \in \mathbb{R}^m$ denote its state and control input at time $t$, respectively. $t\in [0, \infty)$ denotes the continuous-time index.

\begin{figure}[ht]
    \centering
    \includegraphics[width=0.46\textwidth]{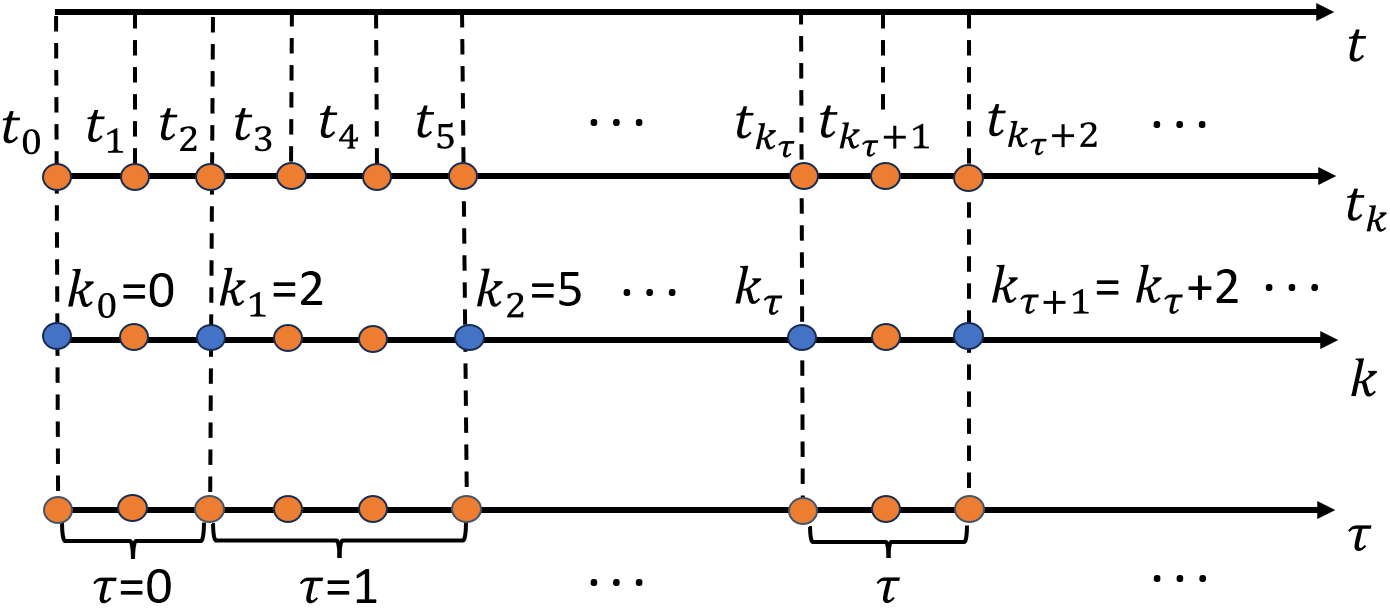}
    \caption{Time indexes, where blue color denotes $k_\tau$.}
    \label{fig:timeindex}
\end{figure}

Suppose the states and control inputs can be obtained from unknown continuous NTVS at certain sampling time instances $t_k \in[0,\infty)$ with $k=0, 1, 2, \cdots$ the index of sampled data points. For notation brevity, one denotes $x_k\coloneqq x_{t_k}$ and $u_k\coloneqq u_{t_k}$ as the $k$-th observed system state and control input, respectively, in the remainder of this manuscript. Then one can partition the observed states-inputs pairs as the following series of data batches
\begin{equation} \label{eq_batch}
    \mathcal{B}_\tau=\{x_k,u_k:  k\in \mathbb{K}_\tau \}, \quad \tau=0,1,2,\cdots,
\end{equation} 
where $$\mathbb{K}_\tau=\{k_\tau,k_\tau+1,k_\tau+2,\cdots,k_{\tau}+\beta_\tau\}$$ denotes the ordered labels set of sampling instances for the $\tau$-th data batch $\mathcal{B}_\tau$ with $\beta_\tau$ positive integers such that $$k_\tau=\sum_{i=0}^{\tau-1} \beta_i,\quad \tau\geq 1, \quad k_0=0.$$ 
It follows that $$k_{\tau+1}=k_\tau+\beta_\tau, \quad\tau=0,1,2,\cdots, $$ which implies the last data in the $\tau$-th data batch is the first data in $(\tau+1)$-th data batch.
For notation simplicity, one defines $\mathcal{B}_\tau^x \coloneqq \{x_k:  k\in \mathbb{K}_\tau\}$, $\mathcal{B}_\tau^u \coloneqq \{u_k:  k\in \mathbb{K}_\tau\}$ in the rest of this manuscript. An illustration of the above indexes is shown in Fig. \ref{fig:timeindex}.

This paper aims to develop an iterative method that approximates the dynamics of an unknown NTVS based on available data batches $\mathcal{B}_\tau$ by a linear time-varying discrete-time system. One way to achieve such a linear approximation is by employing the Koopman operator as in \cite{yeung2019learning,dk2,dk,bevanda2021koopmanizingflows}. Namely, based on the data batch $\mathcal{B}_\tau$, one finds a nonlinear mapping $g(\cdot, \theta_{\tau}): \mathbb{R}^{n} \rightarrow \mathbb{R}^{r}$ parameterized by $\theta_\tau\in\mathbb{R}^q$ \footnote{ Here, $g(\cdot, \theta_{\tau})$ is usually represented by a DNN with a known structure $g$ and an adjustable parameter $\theta_\tau\in\mathbb{R}^q$.} and constant matrices $A_{\tau} \in \mathbb{R}^{r \times r}$, $B_{\tau} \in \mathbb{R}^{r \times m}$, $C_{\tau} \in \mathbb{R}^{n \times r}$  such that for $k\in \mathbb{K}_\tau, k<k_\tau+\beta_\tau$, the following holds approximately,
% \begin{equation}
\begin{align}
    g(x_{k+1},\theta_{\tau}) &= A_{\tau}g(x_{k},\theta_{\tau}) + B_{\tau}u_k, \label{eqq1} \\ x_{k+1} &= C_{\tau}g(x_{k+1},\theta_{\tau}).\label{eqq2}
\end{align} Here, $g(\cdot, \theta_\tau)$, $A_\tau$, $B_\tau$, $C_\tau$ achieved from $\mathcal{B}_\tau$ are put together in the following $\mathcal{K}_{\mathcal{B}_\tau}$:
\begin{equation}\label{eq_dkr}
\mathcal{K}_{\mathcal{B}_\tau} \coloneqq \{g(\cdot, \theta_\tau),A_\tau,B_\tau,C_\tau\},
\end{equation} which is called a deep Koopman representation (DKR) in this manuscript.

Based on the DKR in (\ref{eq_dkr}), one could introduce $\hat{x}_k\in \mathbb{R}^n$ for $k\in \mathbb{K}_\tau, k<k_\tau+\beta_\tau$ as follows:
\begin{align}
    g(\hat x_{k+1},\theta_{\tau}) &= A_{\tau}g(\hat x_{k},\theta_{\tau}) + B_{\tau}\hat u_k, \label{eqqq1} \\ \hat x_{k+1} &= C_{\tau}g(\hat x_{k+1},\theta_{\tau}), \label{eqqq2}
\end{align}
where \begin{equation} \label{eq_initial}   \hat{u}_k=u_k, \forall k\in \mathbb{K}_\tau, k<k_\tau+\beta_\tau, \quad \hat{x}_{k_\tau}=x_{k_\tau}.
\end{equation} This leads to the following linear system
\begin{equation}\label{eqqt0}
\hat{x}_{k+1}=\hat{A}_\tau \hat{x}_{k} +\hat{B}_\tau \hat{u}_{k}, \quad  k\in \mathbb{K}_\tau, k<k_\tau+\beta_\tau,
\end{equation} 
 with $\hat{A}_{\tau}= C_{\tau}A_{\tau}C_{\tau}^\dagger$, $\hat{B}_{\tau}=C_{\tau}B_{\tau}$ and initial conditions in (\ref{eq_initial}). The system \eqref{eqqt0} can be viewed as a linear approximation to the NTVS based on the data batch $\mathcal{B}_\tau$. Note that when for any data batch $\mathcal{B}_\tau$, one has $A_\tau$, $B_\tau$,$ C_\tau$ remain constant for $k\in\mathbb{K}_\tau$. It follows that \eqref{eqqt0} is a linear time-invariant system for $k\in\mathbb{K}_\tau$.
 
\hspace{-0.1cm} To sum up, the \textbf{problem of interest} is to develop an iterative update rule to achieve a DKR in \eqref{eq_dkr} based on data batch $\mathcal{B}_\tau$ in \eqref{eq_batch} such that the linear system  \eqref{eqqt0} is a nice approximation of the unknown NTVS, i.e. $\hat{x}_{k}$ \eqref{eqqt0} is close to $x_{k}$ observed in $\mathcal{B}_\tau$ from the unknown NTVS in the sense that for any given accuracy $\varepsilon\geq 0$, $\hat{u}_k = u_k$ and $\hat{x}_{k_\tau} = x_{k_\tau}$, the estimation error $\parallel\hat{x}_{k}-x_{k}\parallel \leq \varepsilon$.

\section{Main Results}\label{pmethod}
% In this section, we first propose an algorithm to achieve a DKR to approximate an unknown NTVS and then investigate the estimation error between the state from this DKR in \eqref{eqqt0} and the observed state of the unknown NTVS.
This section proposes an algorithm to achieve a deep Koopman representation (DKR) that can approximate an unknown NTVS. We then investigate the estimation error between the state obtained from this DKR, as given in \eqref{eqqt0}, and the observed state of the unknown NTVS.
\subsection{Key Idea}
% Motivated by the methods based on the deep Koopman operator developed in \cite{yeung2019learning,dk2,dk,bevanda2021koopmanizingflows}, in order to achieve a $\theta_\tau^*$ (an optimal $\theta_\tau$ for the DKR), one only needs to solve the following optimization problem based on the data batch $\mathcal{B}_\tau$:
Motivated by deep Koopman operator-based methods developed in \cite{yeung2019learning, dk2, dk, bevanda2021koopmanizingflows}, an optimal $\theta_\tau$ for the deep Koopman representation (DKR), denoted by $\theta_\tau^*$, can be obtained by solving the following optimization problem based on the data batch $\mathcal{B}_\tau$:
\begin{equation} \label{eq_obj0}
   \theta_\tau^*= \arg\min_{\theta_{\tau}\in\mathbb{R}^q} \{ w \bL_1(A_\tau,B_\tau,\theta_\tau) +(1-w)\bL_2(C_\tau,\theta_\tau)\},
\end{equation}
% \begin{equation} \label{eq_L1}
% \begin{aligned}
% \bL_1(\theta_\tau)=\min_{A_\tau\in\mathbb{R}^{r\times r}, B_\tau\in\mathbb{R}^{r\times m}}\bL_{11}(A_{\tau}, B_{\tau}) + \\ \min_{C_{\tau}\in\mathbb{R}^{n\times r}}\bL_{12}(C_\tau),
% \end{aligned}
% \end{equation}
where
\begin{equation} \label{pro1}
\begin{aligned}
\bL_1(A_{\tau}, B_{\tau},\theta_\tau)  = \frac{1}{\beta_{\tau}}\sum_{k=k_\tau}^{k_\tau+\beta_{\tau}-1}\parallel g(x_{k+1}, \theta_{\tau}) 
    -(A_{\tau}g(x_k, \theta_{\tau}) + B_{\tau}u_k)\parallel^2
\end{aligned}
\end{equation}
and
\begin{equation}\label{pro11}
    \bL_2(C_\tau,\theta_\tau)= \frac{1}{\beta_{\tau}} \sum_{k=k_\tau}^{k_\tau+\beta_{\tau}-1}\parallel x_k - C_{\tau}g(x_k, \theta_{\tau})\parallel^2.
\end{equation} 
% come from approximation of (\ref{eqq1}) and (\ref{eqq2}), respectively. $0< w\leq 1$ is a constant which combines the objective of minimization of $\bL_1$ and $\bL_2$. In words, $\bL_{1}$ and $\bL_{2}$ denote the simulation errors in the lifted and original coordinates, respectively.
The objectives of \eqref{pro1} and \eqref{pro11} are to approximate \eqref{eqq1} and \eqref{eqq2}, respectively. Here, $0< w< 1$ is a constant that combines the objective of minimizing $\bL_1$ and $\bL_2$. In simple terms, $\bL_{1}$ and $\bL_{2}$ measure the simulation errors in the lifted and original coordinates, respectively.
% ; and
% \begin{equation}\label{ctrl}
%     \bL_2(\theta_\tau) =  (n-\text{rank}(o(\hat{A}_\tau, \hat{B}_\tau)))^2,
% \end{equation} 
% where $ o(\hat{A}_\tau, \hat{B}_\tau)$ denotes the controllability matrix for the linear system with $\hat{A}_\tau = C_{\tau}A_{\tau}C_{\tau}^\dagger, \hat{B}_\tau=C_{\tau}B_{\tau}$ and $A_\tau, B_\tau, C_\tau$ from minimization of $\bL_{11}$ and $\bL_{12}$; $0< w\leq 1$ is a constant which combines the objective of minimization of $\bL_1$ to achieve a DKR and minimization of $\bL_2$ for the resulted DKR to be controllable.
% \begin{remark}
% In words, $\bL_{11}$ and $\bL_{12}$ denote the simulation errors in the lifted and original coordinates, respectively. $\bL_2$ in \eqref{ctrl} is introduced to make the resulted linear system in \eqref{eqqt0} controllable for the purpose of developing controllers later. $w=1$ means that one does not care about whether the resulted DKR is controllable. 
% \end{remark}

To solve \eqref{eq_obj0}, one needs to rewrite the available data batch and objective functions $\bL_1$ and $\bL_2$ in compact forms.
Toward this end, the following notation is introduced:
\begin{equation}\label{xyudata}
    \begin{aligned}
    \bX_{\tau} &=[x_{k_\tau}, x_{k_\tau+1},\cdots,x_{k_\tau+\beta_{\tau}-1}] \in \mathbb{R}^{n \times \beta_{\tau}},\\
    \bar{\bX}_{\tau} &= [x_{k_\tau+1}, x_{k_\tau+2},\cdots,x_{k_\tau+\beta_{\tau}}]\in \mathbb{R}^{n \times \beta_{\tau}}\nonumber,\\
    \mathbf{U}_{\tau} &=[u_{k_\tau}, u_{k_\tau+1},\cdots,u_{k_\tau+\beta_{\tau}-1}]\in \mathbb{R}^{m \times \beta_{\tau}}.\nonumber
    \end{aligned}
\end{equation}
% \begin{eqnarray}\label{xyudata}
%     \bX_{\tau} \!\!\!&=&\!\!\![x_{k_\tau}, x_{k_\tau+1},\cdots,x_{k_\tau+\beta_{\tau}-1}] \in \mathbb{R}^{n \times \beta_{\tau}},\\
%     \bar{\bX}_{\tau} \!\!\!&=&\!\!\! [x_{k_\tau+1}, x_{k_\tau+2},\cdots,x_{k_\tau+\beta_{\tau}}]\in \mathbb{R}^{n \times \beta_{\tau}}\nonumber,\\
%     \mathbf{U}_{\tau} \!\!\!&=&\!\!\! [u_{k_\tau}, u_{k_\tau+1},\cdots,u_{k_\tau+\beta_{\tau}-1}]\in \mathbb{R}^{m \times \beta_{\tau}}.\nonumber
% \end{eqnarray} 
Then $\bL_{1}$ in (\ref{pro1}) and $\bL_{2}$ in (\ref{pro11}) can be rewritten as
\begin{equation}\label{dktvmins1}
\begin{aligned}
     \bL_{1}= \frac{1}{\beta_{\tau}}\parallel\bar{\bG}_\tau- (A_{\tau}\bG_\tau + B_{\tau}\mathbf{U}_{\tau})\parallel_F^2
\end{aligned}
\end{equation} and
\begin{equation}\label{dktvmins2}
\begin{aligned}
     \bL_{2} = \frac{1}{\beta_{\tau}}\parallel \bX_{\tau} - C_{\tau}\bG_\tau \parallel_F^2, 
\end{aligned}
\end{equation} where
\begin{equation}\label{gstack1}
\begin{aligned}
    \bG_\tau&= [g(x_{k_\tau},\theta_{\tau}),\cdots, g(x_{k_\tau+\beta_{\tau}-1}, \theta_{\tau})] \in \mathbb{R}^{r \times \beta_{\tau}},\\
    \bar{\bG}_\tau&= [g(x_{k_\tau+1}, \theta_{\tau}),\cdots, g(x_{k_\tau+\beta_{\tau}}, \theta_{\tau})] \in \mathbb{R}^{r \times \beta_{\tau}}.
\end{aligned}
\end{equation} 
By minimizing $\bL_{1}$ with respect to $A_\tau, B_\tau$ in \eqref{dktvmins1} and minimizing $\bL_{2}$ regarding $C_\tau$ in \eqref{dktvmins2}, $A_\tau, B_\tau, C_\tau$ can be determined by $\theta_\tau$ as follows:
\begin{align}
    [A_{\tau}^{\theta}, B_{\tau}^{\theta }] &= \bar{\bG}_\tau \begin{bmatrix} \bG_\tau \\ \mathbf{U}_\tau \end{bmatrix}^\dagger \label{lmn},\\
  C_{\tau}^{\theta } &= \bX_{\tau}\bG_\tau^{\dagger}. \label{lmn1}
\end{align}
Replacing $A_\tau$ and $B_\tau$ in \eqref{pro1} by \eqref{lmn} and $C_\tau$ in \eqref{pro11} by \eqref{lmn1}, the objective function in \eqref{eq_obj0} can be reformulated as
\begin{equation}\label{lossf1}
    \begin{aligned}
        \bL(\theta_{\tau})=\frac{1}{\beta_\tau}\!\! \sum_{k=k_\tau}^{k_\tau+\beta_\tau-1} \!\!\parallel\!\! \begin{bmatrix}
        g(x_{k+1},\theta_{\tau}) \\ {x_k}
        \end{bmatrix}\!-\! K_{\tau}^{\theta } \begin{bmatrix}
        g(x_k,\theta_{\tau}) \\ {u_k}
        \end{bmatrix} \!\!\parallel^2,
    \end{aligned}
\end{equation}
with
        \[\ K_{\tau}^{\theta } = \begin{bmatrix}
        A_{\tau}^{\theta } &B_{\tau}^{\theta  }\\ C_{\tau}^{\theta } & \mathbf{0}_{n\times m}
        \end{bmatrix}.\]
% Similarly,  substituting $A_\tau, B_\tau , C_\tau$ in \eqref{lmn} and \eqref{lmn1} into \eqref{ctrl} yields
% \begin{equation}\label{lossf2}
%     \bL_2(\theta_\tau) =  (n-\text{rank}(o(\hat{A}_{\tau}^{\theta_{\tau}},\hat{B}_{\tau}^{\theta_{\tau}})))^2,
% \end{equation}
% where
% \begin{equation}
%     \begin{aligned}
%         \hat{A}_{\tau}^{\theta_{\tau}} &=C_{\tau}^\thetaA_{\tau}^\theta{C_{\tau}^\theta}^\dagger ,\\ \hat{B}_{\tau}^{\theta_{\tau}}&=C_{\tau}^\thetaB_{\tau}^\theta, \nonumber
%     \end{aligned}
% \end{equation}
% with $A_\tau^{\theta_{\tau}}, B_\tau^{\theta_{\tau}}, C_\tau^{\theta_{\tau}}$ given in \eqref{lmn}-\eqref{lmn1}.
% \begin{remark}
% $\bL_{1}(\theta_{\tau})$ in \eqref{lossf1} is a compact form of \eqref{pro1}-\eqref{pro11}.
% \end{remark}

% Here, if one applies the existing deep Koopman operator methods developed in \cite{yeung2019learning,dk2,dk,bevanda2021koopmanizingflows} to achieve the DKR by solving \eqref{eq_obj0} based on each $\mathcal{B}_\tau$ available, there will be two shortcomings. First, one needs to compute the pseudo-inverse in \eqref{lmn} and \eqref{lmn1} repeatedly while solving \eqref{eq_obj0}, which is computationally expensive as $\tau$ increases. Additionally, $\theta_{\tau}$ is required to be initialized for each $\mathcal{B}_{\tau}$, which may be challenging for time-varying systems applications.
Applying the existing deep Koopman operator methods developed in \cite{yeung2019learning, dk2, dk, bevanda2021koopmanizingflows} to achieve the DKR by solving \eqref{eq_obj0} based on each $\mathcal{B}_\tau$ available has two shortcomings. First, computing the pseudo-inverse in \eqref{lmn} and \eqref{lmn1} repeatedly while solving \eqref{eq_obj0} becomes computationally expensive as $\tau$ increases. Second, $\theta_{\tau}$ must be initialized for each $\mathcal{B}_{\tau}$, which can be challenging in time-varying systems applications. To overcome these two limitations, such that one can apply the deep Koopman operator method to approximate the unknown NTVS efficiently, we propose the following method.
\subsection{Algorithm}
Before proceeding, we need the following assumption.
\begin{assumption}\label{asp1}
The matrix $\bG_\tau \in \mathbb{R}^{r\times \beta_\tau}$ in \eqref{gstack1} and $\begin{bmatrix} \bG_\tau \\ \mathbf{U}_{\tau} \end{bmatrix}\in \mathbb{R}^{(r+m)\times \beta_\tau}$ are of full row rank.
\end{assumption}
\begin{remark}
Assumption \ref{asp1} is to ensure the matrices $\bG_\tau \in \mathbb{R}^{r\times \beta_\tau}$ and $\begin{bmatrix} \bG_\tau \\ \mathbf{U}_{\tau} \end{bmatrix}\in \mathbb{R}^{(r+m)\times \beta_\tau}$ invertible and it naturally requires $\beta_{\tau}\geq r+m$.
\end{remark}
\begin{lemma}\label{lemma1}
Given $\mathcal{K}_{\mathcal{B}_\tau}$ in \eqref{eq_dkr}, if Assumption \ref{asp1} holds, then the matrices $A_{\tau+1}^{\theta }, B_{\tau+1}^{\theta }$, $C_{\tau+1}^{\theta }$ can be achieved by
\begin{equation}\label{tau1}
\begin{aligned}
        [A_{\tau+1}^\theta, B_{\tau+1}^\theta] = (\bar{\bG}_{\tau+1} - [A_{\tau}^\theta, B_{\tau}^\theta]\chi_{\tau+1}) \lambda_{\tau} \chi_{\tau+1}^T (\chi_{\tau}\chi_{\tau}^T)^{-1} + [A_{\tau}^\theta, B_{\tau}^\theta],
    \end{aligned}
\end{equation}
\begin{equation}\label{tau2}
\begin{aligned}
        C_{\tau+1}^{\theta } = (\bX_{\tau+1} -C_\tau^\theta\bG_{\tau+1})\bar{\lambda}_{\tau}\bG_{\tau+1}^T(\bG_\tau\bG_\tau^T)^{-1} + C_\tau^\theta,
    \end{aligned}
\end{equation}
where \begin{equation}
    \begin{aligned}
        \chi_{\tau} &= \begin{bmatrix}
              \bG_\tau\\  
              \mathbf{U}_{\tau}
         \end{bmatrix} \in \mathbb{R}^{(r+m) \times \beta_\tau}, \\
         \lambda_{\tau} &=(\bI_{\beta_{\tau+1}}+\chi_{\tau+1}^T (\chi_{\tau}\chi_{\tau}^T)^{-1}\chi_{\tau+1})^{-1}\in\mathbb{R}^{\beta_{\tau+1} \times \beta_{\tau+1}},\\
         \bar{\lambda}_{\tau} &=(\bI_{\beta_{\tau+1}}+\bG_{\tau+1}^T (\bG_\tau\bG_\tau^T)^{-1}\bG_{\tau+1})^{-1}\in \mathbb{R}^{\beta_{\tau+1} \times \beta_{\tau+1}}. \nonumber
    \end{aligned}
\end{equation}
Here, $\bG_{\tau+1} \in \mathbb{R}^{r\times \beta_{\tau+1}}$ and $\bar{\bG}_{\tau+1} \in \mathbb{R}^{r\times \beta_{\tau+1}}$ are defined in \eqref{gstack1}.
\end{lemma}
The proof of Lemma~\ref{lemma1} will be given in Appendix.
% The proof of Lemma~\ref{lemma1} is given in the Appendix.

To initialize the algorithm, one first needs to build the DNN $g(\cdot, \theta_{\tau}): \mathbb{R}^{n} \rightarrow \mathbb{R}^{r}$ with non-zero $\theta_{\tau}\in\mathbb{R}^q$. Then the matrices $A_{\tau}^\theta\in\mathbb{R}^{r\times r}, B_{\tau}^\theta\in\mathbb{R}^{r\times m}$ and $C_{\tau}^\theta\in\mathbb{R}^{n\times r}$ can be found by solving \eqref{lmn} and \eqref{lmn1}, respectively, based on $\mathcal{B}_{\tau}$. When the data batch $\mathcal{B}_{\tau+1}$ becomes available, one can directly update the $\bG_{\tau+1}$ and $\bar{\bG}_{\tau+1}$ by following \eqref{gstack1}, which leads to the $\chi_{\tau+1}$, $\lambda_\tau$ and $\bar{\lambda}_\tau$ in Lemma~\ref{lemma1}. Thus one can update $A_{\tau}^\theta$, $B_{\tau}^\theta$ and $C_{\tau}^\theta$ efficiently by computing \eqref{tau1}-\eqref{tau2} instead of solving \eqref{lmn}-\eqref{lmn1} repeatedly as a consequence of applying Lemma~\ref{lemma1}. Finally, an optimal $\theta_{\tau+1}^*$ is obtained by solving \eqref{eq_obj0} with $A_{\tau+1}^\theta, B_{\tau+1}^\theta$ subject to \eqref{tau1} and $C_{\tau+1}^\theta$ satisfying \eqref{tau2}  based on $\mathcal{B}_{\tau+1}$ .

To sum up, we have the following algorithm, which is referred to as deep Koopman learning for time-varying systems (DKTV) in the remainder of this manuscript and its pseudocode is provided in Appendix.
\begin{enumerate}
    \item Initialization: Build $g(\cdot, \theta_{\tau}): \mathbb{R}^{n} \rightarrow \mathbb{R}^{r}$ with $\theta_{\tau}\in\mathbb{R}^q, \theta_{\tau}\neq \mathbf{0}_q$; initialize $A_{\tau}^\theta, B_{\tau}^\theta$ and $C_{\tau}^\theta$ by solving \eqref{lmn} and \eqref{lmn1}, respectively, based on $\mathcal{B}_{\tau}$.
    \item When $\mathcal{B}_{\tau+1}$ becomes available, update $A_{\tau}^\theta$, $B_{\tau}^\theta$ and $C_{\tau}^\theta$ according to \eqref{tau1} and \eqref{tau2}, respectively.
    \item Solve \eqref{eq_obj0} to find $\theta_{\tau+1}^*, A_{\tau+1}^{\theta^*}, B_{\tau+1}^{\theta^*}$ and $C_{\tau+1}^{\theta^*}$.
    \item Repeat steps $2$-$3$ as the unknown NTVS evolves.
\end{enumerate}

\subsection{Analysis of the Approximation Error}
With a slight abuse of notation, suppose one observes the latest $\mathcal{B}_\tau$ with $\{x_{k-1}, u_{k-1}\}$ its latest data point (i.e., $x_{k-1}\coloneqq x_{k_\tau+\beta_\tau}$, $u_{k-1}\coloneqq u_{k_\tau+\beta_\tau}$), in this subsection, we investigate the estimation errors induced by the proposed algorithm described as: 
\begin{equation}\label{eq_error}
    e_k = \hat{x}_k - x_k,
\end{equation}
where $\hat{x}_k\in\mathbb{R}^n$ denotes the estimated state achieved by the proposed method and $x_k\in\mathbb{R}^n$ denotes the state of the unknown NTVS evolving from $x_{k-1}\in\mathbb{R}^n$ and $u_{k-1}\in\mathbb{R}^m$. Here, to analyze the \eqref{eq_error} with respect to $\theta_\tau\in\mathbb{R}^q$, one rewrites \eqref{eqqq1}-\eqref{eqqq2} as:
\begin{equation}\label{xhat}
    \hat{x}_{k} = C_{\tau}(A_{\tau}g(\hat{x}_{k-1},\theta_\tau)+B_\tau \hat{u}_{k-1}),
\end{equation}
with condition \eqref{eq_initial} hold.

Before we present the results, the following concepts about operators are introduced.
\subsubsection{Preliminaries}
\begin{definition}[Koopman operator with control]\label{defi1}
Consider the Hilbert space $\mathcal{F}$ and discrete-time nonlinear time-varying system $x_{k+1}=f(x_k,u_k,k)$ starting from time $k_0$, where $k\in\mathbb{Z}$ denotes its time index; $x_k\in\mathcal{M}$ and $u_k\in\mathcal{U}$ denote the system state and control input at time $k$, respectively; $\mathcal{M}\subseteq\mathbb{R}^n$ and $ \mathcal{U}\subseteq\mathbb{R}^m$ denote the state space and control input space, respectively. Then the dynamics of the states of the augmented control system 
$z_k,z_{k+1}$ is described by
\[z_{k+1} = F(z_k, k) \coloneqq \matt{f(x_k, {\boldsymbol u}(0), k)\\ \mathcal{S}{\boldsymbol u}},
\]
where $z_k = \matt{x_k\\ \boldsymbol{u}}$ denotes the augmented control system states; $\mathcal{S}\boldsymbol u$ denotes the left shifting of the control sequence $\boldsymbol u$, i.e., $(\mathcal{S}{\boldsymbol u})(i) = {\boldsymbol u}(i + 1)$ and ${\boldsymbol u}(i)$ denotes the $i$th element of the control sequence $\boldsymbol u$.

We then extend the definition of the Koopman operator from \cite{yi2021equivalence} to discrete-time NTVS. Let operator $\mathcal{K}^{(k_0,k)}: \mathcal{F} \rightarrow \mathcal{F}$ act on functions of state space $\phi(\cdot)\in\mathcal{F}$ with $\phi(\cdot): \mathcal{M}\times\mathcal{U}\rightarrow \mathbb{C}$ and defined with two parameters $(k_0,k)$ as
\begin{equation}\label{ko1}
    \mathcal{K}^{(k_0,k)} [\phi(z_k)] = \phi(F(z_k,k)).\nonumber
\end{equation}
\end{definition}
% Note that the Koopman operator maps functions of state space to functions of state space instead of states to states \cite{korda2018linear}.

\begin{definition}[$L_2$-projection~\cite{korda2018convergence}]\label{defi2}
For brevity, we recall the definitions of $\mathcal{F}$, $\mathcal{M}$, and $\phi(\cdot)$ from Definition \ref{defi1}. Let $\nu\in\mathbb{R}$ be the positive measurement on $\mathcal{M}$ and assume that $\mathcal{F} = L_2(\nu)$. Given a set of linearly independent $\phi_i\in\mathcal{F}$ with $i = 1, 2, \cdots, r$, and define 
\begin{align*}
    \mathcal{F}_r&\coloneqq\text{span}\{\phi_1, \phi_2,\cdots, \phi_r\},\\
    \Phi_r &= [\phi_1, \phi_2,\cdots, \phi_r]^T.
\end{align*} 
Then for all nonzero $c\in\mathbb{R}^r$, the $L_2(\nu)$ projection of a function $\psi\in L_2(\nu)$ on $\mathcal{F}_{r}$ is defined as
\begin{equation}
\begin{aligned}
P_{r}^{\nu}\psi &= \arg\min_{f\in\mathcal{F}_r} \parallel f-\psi\parallel_{L_2(\nu)} \\&= \arg\min_{f\in\mathcal{F}_r}\int_{\mathcal{M}}\parallel f-\psi \parallel^2 d\nu \\ & =\arg\min_{c\in\mathbb{R}^r}\int_{\mathcal{M}} \parallel c^\top\Phi_r - \psi\parallel^2 d\nu .\nonumber
\end{aligned}
\end{equation}
\end{definition}
\begin{definition}[Bounded operator]
Given a Hilbert space $\mathcal{F}$, an operator $\mathcal{O}: \mathcal{F} \rightarrow \mathcal{F}$ is bounded on $\mathcal{F}$ if
$$\parallel\mathcal{O}\parallel \coloneqq \sup_{f\in\mathcal{F},\parallel f\parallel=1} \parallel\mathcal{O}f\parallel<\infty.$$
\end{definition}
$\parallel\mathcal{O}\parallel$ is referred as the operator norm of $\mathcal{O}$ in the remainder of this manuscript.
\begin{definition}[Strong Convergence of operator]
A sequence of bounded operators $\mathcal{O}_i: \mathcal{F} \rightarrow \mathcal{F}$ defined on a Hilbert space $\mathcal{F}$ converges strongly (or in the strong operator topology) to an operator $\mathcal{O}: \mathcal{F} \rightarrow \mathcal{F}$ if \[\lim_{i\rightarrow\infty}\parallel \mathcal{O}_i f - \mathcal{O} f \parallel = 0\]
for all $f\in\mathcal{F}$.
\end{definition}

\subsubsection{Analysis}
We make the following assumptions to show that the error $e_k$ in \eqref{eq_error} is bounded.
\begin{assumption}\label{asp2}
For any $\mathcal{B}_\tau$ in \eqref{eq_batch}, let $\{x_i,u_i\}$ denote its $i$th data point observed at time $t_i$. The observation interval $\Delta t = t_{i+1} - t_i$ is sufficiently small such that for some $\mu_x \geq 0, \mu_u\geq 0$, $\parallel x_{i+1} - x_{i} \parallel \leq \mu_x < \infty$ and $\parallel u_{i+1} - u_i \parallel \leq \mu_u < \infty$. 
\end{assumption}
\begin{assumption}\label{asp3}
The deep neural network observable function $g(\cdot, \theta_\tau)$ is Lipschitz continuous on the system state space with Lipschitz constant $\mu_g$.
\end{assumption}
Consider a DNN observable function $g(\cdot,\theta_\tau): \mathbb{R}^n \rightarrow\mathbb{R}^r$, let $$\bar{\psi}^{h} = [\bar \psi_1^{h}, \bar \psi_2^{h},\cdots, \bar \psi_{n_h}^{h}]^T \in \mathbb{R}^{n_h}$$ be its last hidden layer and $\bar \psi^{o}= [\bar \psi_1^{o}, \bar \psi_2^{o},\cdots, \bar \psi_{r}^{o}]^T \in\mathbb{R}^{r}$ denotes its output layer, where $\bar\psi_i^h:\mathbb{R}\rightarrow \mathbb{R}$ and $\bar\psi_i^o:\mathbb{R}\rightarrow \mathbb{R}$ are generally nonlinear function chosen by user. Note that here $n_h$ and $r$ denote the number of nodes of $\bar{\psi}^{h}$ and $\bar{\psi}^{o}$, respectively.

\begin{assumption}\label{asp4}
(1): Given a Hilbert space $\mathcal{F}$, the Koopman operator $\mathcal{K}$ is bounded and continuous on $\mathcal{F}$; (2): $\bar{\psi}^{h}$ are selected from the orthonormal basis of $\mathcal{F}$, i.e., $[\bar{\psi}_1^{h}, \bar{\psi}_2^{h}, \cdots, \bar{\psi}_\infty^{h}]^T$ is an orthonormal basis of $\mathcal{F}$.
\end{assumption}
\begin{remark}
 Since the DNN observable function $g(x, \theta_\tau)$ is with known structure, one can make Assumption \ref{asp4} hold by choosing proper activation functions for DNN's different layers. We refer to \cite{williams2015data} for more details about candidate functions like radial basis functions.
\end{remark}
\begin{lemma}\label{lemma2}
Let $\mu\in\mathbb{R}$ be the empirical measurements associated with the observed states $x_k\in\mathbb{R}^n$. If Assumption \ref{asp4} holds, then the operator $P_{n_h}^\mu$ converges strongly to the identity operator $I$ as $n_h$ goes to infinity.
\end{lemma}
% Proof of Lemma~\ref{lemma2} is referred to in Appendix.
Proof of Lemma~\ref{lemma2} is referred to in Appendix.

\begin{lemma}\label{lemma3}
Let $P \coloneqq \begin{bmatrix}
    P_{n_h}^{\mu_1} & \mathbf{0}_{n_h\times m}\\ \mathbf{0}_{m\times n_h} & P_m^{\mu_2}
\end{bmatrix}$ with $\mu_1\in\mathbb{R}, \mu_2\in\mathbb{R}$ the empirical measurements associated to the observed system state $x_k\in\mathbb{R}^n$ and control input $u_k\in\mathbb{R}^m$, respectively. If Assumption \ref{asp4} holds, then the approximated sequence of operators $\mathcal{K}_DP=P\mathcal{K}P$ converges to Koopman operator $\mathcal{K}$ strongly as $n_h$ goes to infinity.
\end{lemma}
% Proof of Lemma~\ref{lemma3} is given in Appendix.
Proof of Lemma~\ref{lemma3} is given in Appendix.
\begin{remark}
Lemma~\ref{lemma3} shows the convergence condition of the existing \emph{DKO} method regarding the DNN observable function $g(\cdot, \theta_\tau)$, i.e., $n_h\rightarrow\infty$. It replaces the convergence condition of the EDMD method \cite{korda2018convergence}, i.e., $r\rightarrow\infty$. This conclusion is used in the present work to reduce $\beta_\tau \geq r+m$ so that one can track the unknown NTVS based on $\mathcal{B}_\tau$ with small batch size.
\end{remark}
\begin{thm}\label{mainth}
Let $g(x,\theta_\tau)$ be a DNN with its last hidden layer containing $n_h$ nodes. If Assumptions \ref{asp1}-\ref{asp4} hold, then the estimation error $e_k$ in \eqref{eq_error} is bounded by \begin{equation}
\begin{aligned}
    \lim_{n_h\rightarrow\infty} \sup\parallel e_k\parallel =  (\parallel C_\tau A_\tau \parallel \mu_g + 1) \mu_x + \parallel C_\tau B_\tau \parallel  \mu_u + \max_{\bar{x}\in\mathcal{B}_\tau^x}\parallel \bar{x} - C_\tau g(\bar{x},\theta_\tau) \parallel.\nonumber
\end{aligned}
\end{equation}
\end{thm}
% The proof of Theorem \ref{mainth} is given in Appendix.
The proof of Theorem \ref{mainth} is given in Appendix.
\begin{remark}
Theorem \ref{mainth} says that with certain assumptions hold, as $n_h \rightarrow\infty$, the estimation error of the proposed method depends on the minimization performance of \eqref{dktvmins2} and $\mu_x, \mu_u$ defined in Assumption \ref{asp2}. One way to further reduce the estimation error is to decrease the sampling interval.
\end{remark}
Since one cannot construct a DNN with infinite parameters in practice, we introduce the following corollary based on Theorem \ref{mainth}, for which we need to make the following assumption.
\begin{assumption}\label{asp5}
For any $A_\tau\in\mathbb{R}^{r\times r}$, $\parallel A_\tau \parallel < 1$.
\end{assumption}
\begin{remark}
According to Lemma~\ref{lemma1}, the DNN observable function determines $\parallel A_\tau \parallel$. To make Assumption \ref{asp5} hold, one can add an extra loss function to impose the maximum value of $\parallel A_\tau \parallel$ in \eqref{eq_obj0}. 
\end{remark}
\begin{corollary}\label{cor1}
Recall $e_k$ in \eqref{eq_error}. If Assumptions \ref{asp1}-\ref{asp3} and \ref{asp5} hold, then the upper bound of $\parallel e_k \parallel$ is determined by the minimization performance of \eqref{lossf1} and $\mu_x, \mu_u$ as $k$ goes to infinity.
\end{corollary}
% The proof of Corollary \ref{cor1} is given in Appendix.
The proof of Corollary \ref{cor1} is given in Appendix.

\section{Numerical Simulations}\label{results}
In this section, we first test the proposed algorithm in two examples: one is a simple nonlinear time-varying system (NTVS), and the other is a quadcopter example. In both examples, the system state input data is observed with a fixed time interval of $0.1$s, and $\beta_\tau\equiv \beta$, where $\beta \geq r+m$ is an arbitrary positive integer, and $\tau = \lceil \frac{k-\beta_0}{\beta} \rceil$, where $k\in\mathbb{K}_\tau$. Subsequently, a motivating example is presented to illustrate the effectiveness of the proposed approach in an optimal design setting.

\subsection{A Simple NTVS}
% For discussion, let $x_k\in\mathbb{R}^2$ denote the true system states observed from the unknown NTVS; $\tilde{x}_k \in\mathbb{R}^2$ denotes the estimated states generated by time-varying DMD (TVDMD) method from \cite{zhang2019online}; $\hat{x}_k\in\mathbb{R}^2$ in \eqref{xhat} is the estimated states from the proposed method (DKTV). Accordingly, $\tilde{e}_k = \parallel \tilde{x}_k - x_k\parallel$ and $e_k= \parallel \hat{x}_k - x_k\parallel$ denote the estimation errors induced by TVDMD and DKTV methods, respectively.

Consider the following dynamical system:
\begin{equation}\label{SNTVS}
\dot{x}_t = M_t\cos(x_t),
\end{equation}
where $x_t\in \mathbb{R}^2$ and $M_t$ is a time-varying matrix given by
\begin{equation}\label{TVA}
    M_t = \matt{0 &(1+\gamma t) \\ -(1+\gamma t) &0},
\end{equation}
where $\gamma$ is a constant determining how fast the dynamics change. For more details of the dynamics, see \cite{zhang2019online}.

\begin{figure}[htp]
     \centering
     \begin{subfigure}[b]{0.44\textwidth}
         \centering
         \includegraphics[width=\textwidth]{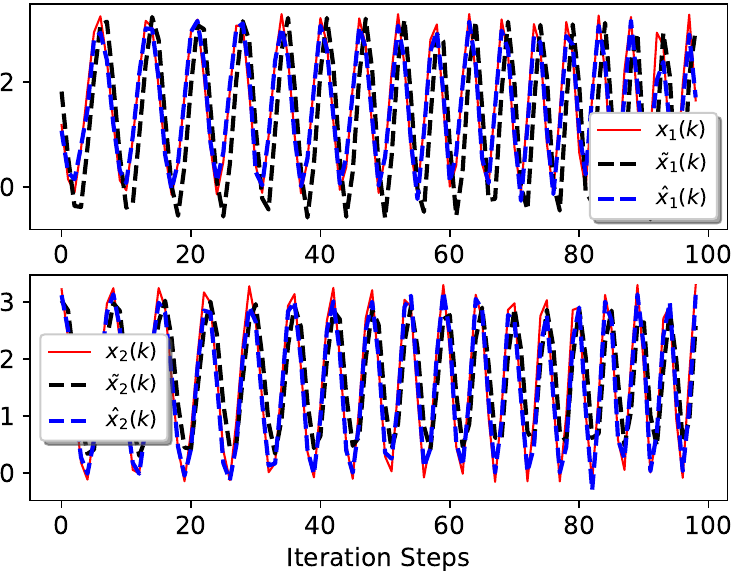}
         \caption{$\gamma=0.8$}
         \label{fig:y equals x}
     \end{subfigure}
     % \hfill
     ~
     \begin{subfigure}[b]{0.44\textwidth}
         \centering
         \includegraphics[width=\textwidth]{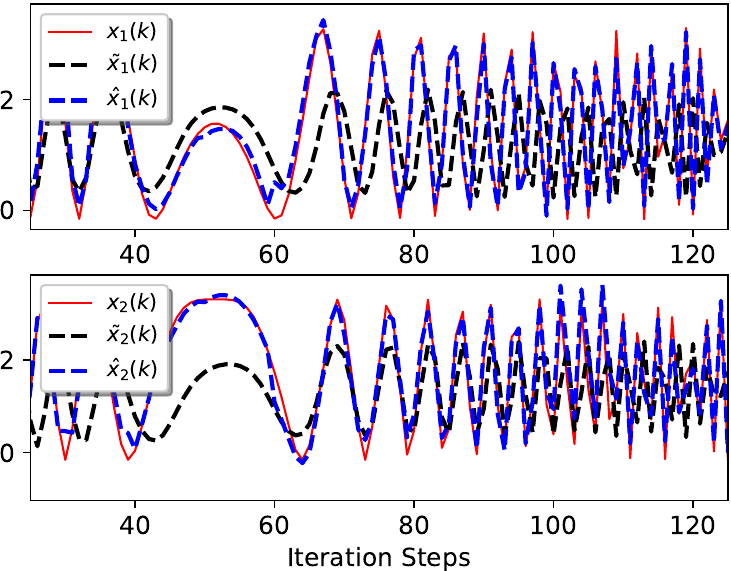}
         \caption{$\gamma=6$}
         \label{fig:three sin x}
     \end{subfigure}
        \caption{System states trajectories: TVDMD (black) vs. DKTV (blue).}
        \label{fig:traj1}
\end{figure}

\begin{figure}[ht]
    \centering
    \includegraphics[width=0.44\textwidth]{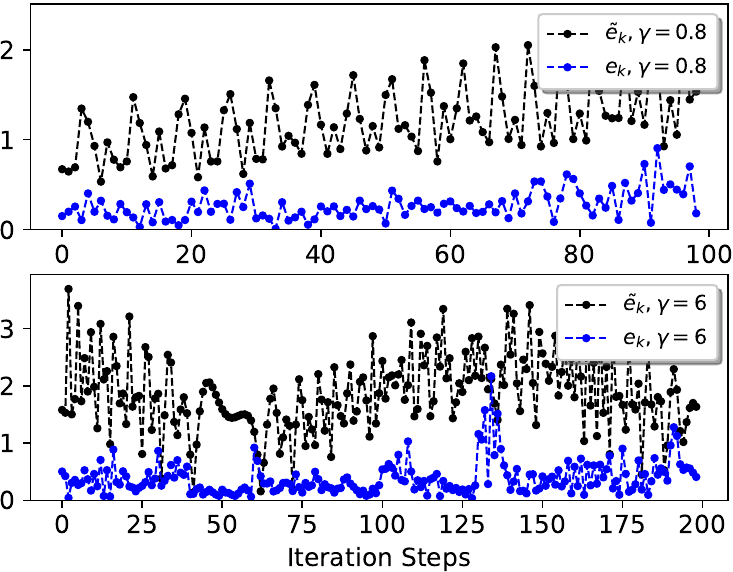}
    \caption{System states estimation errors: TVDMD (black) vs. DKTV (blue).}\label{fig:tracking}
\end{figure}

\noindent{\bf Experiment Setup.} The DNN observable function ($g(\cdot,\theta): \mathbb{R}^2 \rightarrow \mathbb{R}^6$) is built with a hidden layer \emph{ReLu()} that contains $32$ nodes and an output layer \emph{Relu()} consisting of $6$ nodes. The DNN training process is implemented by choosing optimizer \emph{Adam} \cite{kingma2014adam} with \emph{learning rate = 1e-3} and \emph{weight decay rate = 1e-4}. Then $\mathcal{B}_\tau$ is observed with $\beta = 10$ starting from $x_0= [1,0]^T$. 
\begin{remark}
    The activation function $ReLu()$ is selected for this example because by appropriately selecting $x_0$ and $\gamma$, one can ensure that $x_t\geq0$. This choice results in $ReLu(x) = x$ which belongs to the category of linear radial basis functions \cite{schoenberg1938metric}.
\end{remark}
\noindent {\bf Results Analysis.} We test the proposed method in the NTVS in \eqref{SNTVS} for both slow and fast-changing scenarios setting $\gamma = 0.8$ and $\gamma=6$, respectively. The trajectories estimated by time-varying DMD(TVDMD) and the proposed method are shown in Figs. \ref{fig:traj1}, and the estimation errors of both methods are exhibited in Fig. \ref{fig:tracking}. Here, we let $x_k\in\mathbb{R}^2$ denote the true system states observed from the unknown NTVS. Let $\tilde{x}_k \in\mathbb{R}^2$ denote the estimated states generated by the TVDMD method from \cite{zhang2019online}. Let $\hat{x}_k\in\mathbb{R}^2$ be the estimated states from the proposed method (DKTV), as given in \eqref{xhat}. Consequently, $\tilde{e}_k = \parallel \tilde{x}_k - x_k\parallel$ and $e_k= \parallel \hat{x}_k - x_k\parallel$ denote the estimation errors induced by the TVDMD and DKTV methods, respectively.
It can be observed that when the NTVS varies slowly, both methods can capture the dynamics of the NTVS with reasonable accuracy. In situations where the NTVS varies rapidly, the estimation errors of the TVDMD method increase correspondingly. On the contrary, the proposed method can achieve consistent performance, benefiting from its DNN observable function.

% ================================================
% System ID
% ================================================
\subsection{System States Prediction for Single Quadcopter}
In this example, we implement the proposed algorithm to predict the system states of a quadcopter with a time-varying disturbance applied to its body. The dynamics of the quadcopter is described as
{
\begin{equation}\label{quaddyn}
\begin{aligned}
\begin{pmatrix}
\dot{p_n}\\ \dot{p_e} \\ \dot{z}\end{pmatrix} &=\begin{pmatrix}
c_\theta c_\psi & s_\phi s_\theta c_\psi - c_\phi s_\psi & c_\phi s_\theta c_\psi + s_\phi s_\psi \\ c_\theta c_\psi & s_\phi s_\theta s_\psi + c_\phi c_\psi & c_\phi s_\theta s_\psi - s_\phi c_\psi \\ s_\theta & -s_\phi c_\theta & -c_\phi c \theta\end{pmatrix} \begin{pmatrix}
\dot{x}\\ \dot{y} \\ \dot{z}
\end{pmatrix}, \\
\begin{pmatrix}
\dot{\phi}\\ \dot{\theta}\\ \dot{\psi}\end{pmatrix}&= \begin{pmatrix}
1 & s_\phi t_\theta & c_\phi t_\theta\\
0 & c_\phi & -s_\phi\\
0 &\frac{s_\phi}{c_\theta} & \frac{c_\phi}{s_\theta}
\end{pmatrix}\begin{pmatrix} p\\q\\r \end{pmatrix},\\
\begin{pmatrix}
\ddot{x}\\ \ddot{y}\\ \ddot{z}
\end{pmatrix} &= \begin{pmatrix}
r\dot{y}-q\dot{z}\\p\dot{z}-r\dot{x}\\
q\dot{x}-p\dot{y}\\\end{pmatrix} +\frac{1}{m} (\begin{pmatrix}
f_x\\f_y\\f_z \end{pmatrix}+w_t),\\
\begin{pmatrix}
\dot{p}\\ \dot{q}\\ \dot{r}
\end{pmatrix} &= \begin{pmatrix}
\frac{J_y-J_z}{J_x}qr\\\frac{J_z-J_x}{J_y}pr\\
\frac{J_x-J_y}{J_z}pq\\\end{pmatrix} + \begin{pmatrix}
\frac{1}{J_x}\tau_{\phi}\\\frac{1}{J_y}\tau_{\theta}\\
\frac{1}{J_z}\tau_{\psi}\\\end{pmatrix}, \nonumber
\end{aligned}
\end{equation}
}
\hspace{-0.1cm}where $s_\theta$, $s_\phi$, $s_\psi$, $c_\theta$, $c_\phi$, $c_\psi$, and $t_\theta$ denote $\sin(\theta)$, $\sin(\phi)$, $\sin(\psi)$, $\cos(\theta)$, $\cos(\phi)$, $\cos(\psi)$, and $\tan(\theta)$, respectively. $p_n$, $p_e$, and $z$ denote the inertial north position, the inertial east position, and the altitude, respectively. $\dot{x}$, $\dot{y}$, and $\dot{z}$ denote the velocity along the x-axis, y-axis, and z-axis, respectively. $\phi$, $\theta$, and $\psi$ denote the roll angle, pitch angle, and yaw angle, respectively, and $p$, $q$, and $r$ denote the roll rate, pitch rate, and yaw rate, respectively. $f_x$, $f_y$, and $f_z$ denote the total force applied to the quadcopter along the x-axis, y-axis, and z-axis, respectively, and $w_t\in\mathbb{R}^3$ denotes the time-varying disturbance. $\tau_{\phi}$, $\tau_\theta$ and $\tau_\psi$ denote the torque of roll, pitch, and yaw, respectively, while $J_x$, $J_y$, and $J_z$ denote the inertia along the x-axis, y-axis, and z-axis, respectively. For further details on the dynamics of the quadcopter, see \cite{beard2008quadrotor}.

\noindent{\bf Experiment Setup.} In order to compare the performance of the proposed algorithm with that of the single neural network method, we let $x_k \in \mathbb{R}^{12}$ denote the true state of the system, $\bar{x}_k\in \mathbb{R}^{12}$ and $\hat{x}_k\in \mathbb{R}^{12}$ denote the estimated states of the single DNN and the proposed method, respectively, and $\bar{e}_k = \parallel \bar{x}_k - x_k\parallel$ and $e_k= \parallel \hat{x}_k - x_k\parallel$ denote the estimation errors of the single DNN and DKTV methods, respectively. The data batch $\mathcal{B}_\tau$ is collected with $\beta = 30$ manually controlling the quadcopter from position $(0,0,0)$ (meter) to $(1,2,3)$ (meter) with the time-varying disturbance force $w_t\in\mathbb{R}^3$ applied on the quadcopter. $w_t$ is generated from the standard normal distribution. For the DKTV method, the DNN observable function $g(\cdot, \theta): \mathbb{R}^{12} \rightarrow \mathbb{R}^{16}$ is built with one hidden layer \emph{Gaussian function} containing $64$ nodes and one output layer \emph{Relu()} consisting of 16 nodes. For the single DNN method, DNN $N(\cdot, \nu):  \mathbb{R}^{16} \rightarrow \mathbb{R}^{12}$ is built with the same structure as $g(\cdot, \theta)$. Then the DNN training process of both methods is implemented by choosing optimizer \emph{Adam} \cite{kingma2014adam} with \emph{learning rate = 1e-3} and \emph{weight decay rate = 1e-4}. The optimal vector $\nu\in\mathbb{R}^p$ is found by
\begin{equation}\label{loss_sdnn}
    \nu^* = \arg\min_{\nu\in\mathbb{R}^p} L(\nu) = \arg\min_{\nu\in\mathbb{R}^p}\parallel N(x_k,u_k, \nu) -x_{k+1} \parallel,
\end{equation}
which leads to $\bar{x}_{k+1} = N(x_k,u_k, \nu^*)$.

\begin{figure}[ht]
    \centering
    \includegraphics[width=0.44\textwidth]{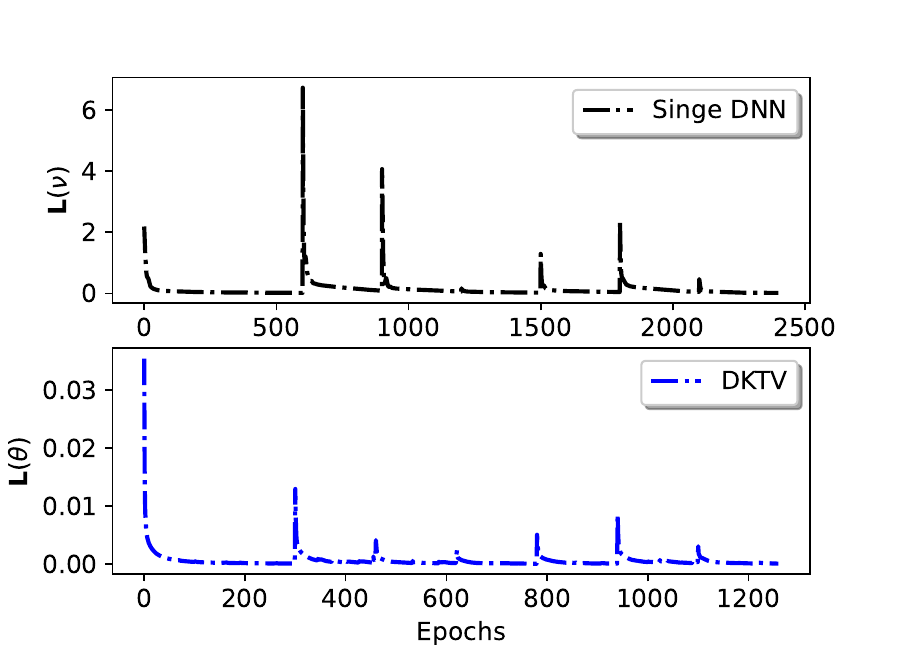}
    \caption{Training process: DNN (black) vs. DKTV (blue).}
    \label{fig:train}
\end{figure}

\begin{figure}[ht]
    \centering
    \includegraphics[width=0.42\textwidth]{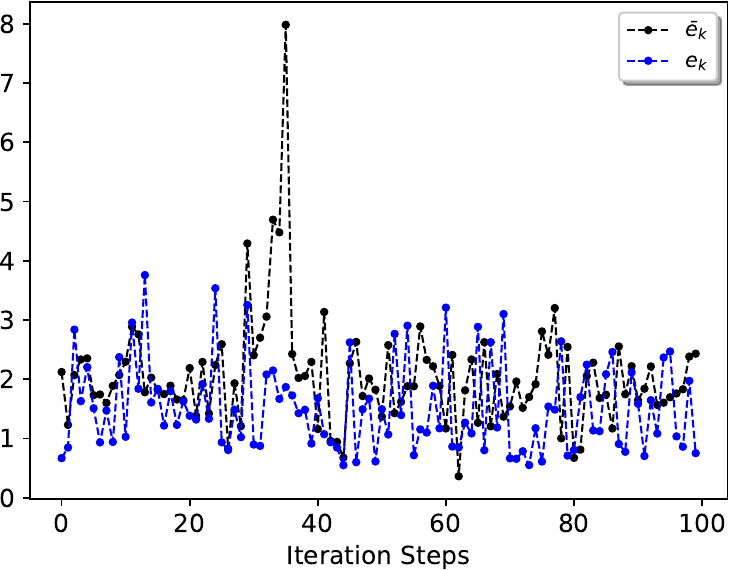}
    \caption{System states estimation error: DNN (black) vs. DKTV(blue).}
    \label{fig:predqua}
\end{figure}

\begin{figure}[ht]
    \centering
    \includegraphics[width=0.42\textwidth]{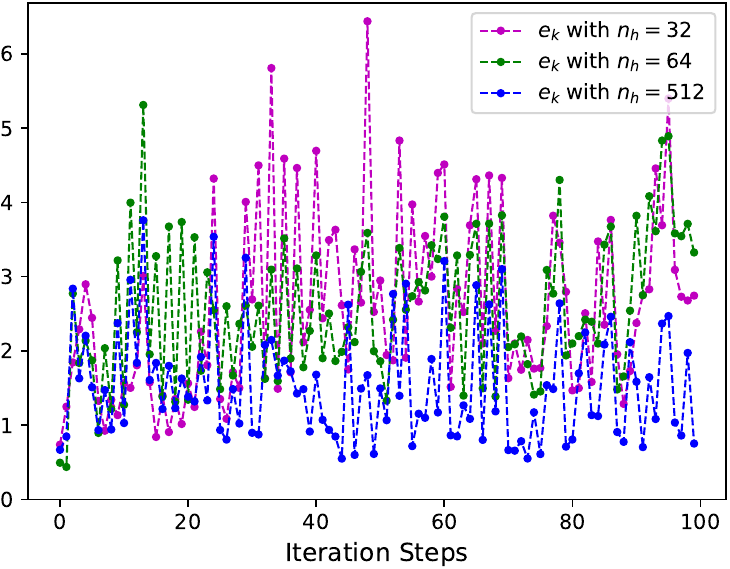}
    \caption{The estimation errors of DKTV with different DNN observable functions.}
    \label{fig:errornorm}
\end{figure}

\noindent {\bf Results Analysis.} We first show the DNN training process of both methods with $\tau=1,\cdots,7$ in Fig. \ref{fig:train}, where one epoch denotes one forward and one backward pass through the DNN and $L(\nu), L(\theta)$ are defined in \eqref{loss_sdnn} and \eqref{lossf1}, respectively. As shown, DKTV needs fewer training epochs to converge and maintains a much lower loss value during the training process than the single DNN method, as expected due to its double minimization of \eqref{dktvmins1}-\eqref{dktvmins2}.
Then in Fig. \ref{fig:predqua}, we demonstrate the algorithm performance by showing the system states estimation errors of both methods. As is shown, the proposed algorithm can provide better prediction performance compared with the Single DNN method.
Finally, in Fig. \ref{fig:errornorm}, Theorem \ref{mainth} is validated by showing that the estimation errors of the proposed algorithm are reduced by increasing the number of nodes of its hidden layer.
\subsection{Optimal Control for Time-varying Cartpole}
In this subsection, we apply the proposed method to design an optimal controller for the classical cartpole system, as formulated in \cite{barto1983neuronlike}. The system dynamics are described as follows:
\begin{align}\label{Cartpoledyn}
    \ddot{x}_t &= \frac{F_t+ml (\dot{\bar{\theta}}_t^2\sin{\bar{\theta}_t} - \ddot{\bar{\theta}}_t \cos{\bar{\theta}_t}) - \mu_{t}^c \sgn{(\dot{x}_t})}{m_c+m} \\\nonumber
    \ddot{\bar{\theta}}_t &= \frac{\cos{\theta_t}[-F_t - ml \dot{\bar{\theta}}_t^2\sin{\theta_t} +\mu_{t}^c \sgn{(\dot{x}_t})]/(m_c+m)}{l[\frac{4}{3} - (m \cos^2{\bar{\theta}_t})/(m_c+m)]} \\ &+ \frac{g\sin{\theta_t}-\mu_p \dot{\bar{\theta}}/{ml}}{l[\frac{4}{3} - (m \cos^2{\bar{\theta}_t})/(m_c+m)]}, \nonumber
\end{align}
where $\dot{\mu}_{t}^c=0.3\cos(t)$ with $\mu_0^c = 0.0005$ is the \emph{time-varying} coefficient of friction of cart on track; $x_t$ denotes the distance of the cartpole moves from the initial position; $\bar{\theta}_t$ denotes the angle from the up position; $\dot{x}_t,\dot{\bar\theta}_t$ denote the x-axis velocity and the angular velocity respectively; $F_t$ denotes the continuous control input applied to the center of the mass of the cart at time step $t$; $g=-9.8m/s^2$ is the gravity acceleration; $m_c=1.0kg, m=0.1kg$ are the mass of the cart and the pole, respectively; $l=0.5m$ is the length of the pole; $\mu_p = 0.000002$ is the coefficient of the friction of the pole on cart. 

In this example, the coefficient of friction of the cart on the track, denoted $\mu_c(t)$, is made time-varying during the simulation. A deep neural network $g(\cdot,\theta): \mathbb{R}^4 \rightarrow \mathbb{R}^6$ is first constructed. The data batch $\mathcal{B}_\tau$ is generated by observing the dynamics in \eqref{Cartpoledyn} with batch size $\beta = 12$ and fixed time interval $t_{k_\tau+1}-t_{k_\tau} \equiv 0.1s$. The learned dynamics are subsequently employed to design a model predictive control (MPC) controller aimed at stabilizing the cartpole system around the upright equilibrium, defined by $\bar{\theta} = 0, \dot{\bar{\theta}} = 0$.
The optimization problem is formulated as
\begin{mini}|s|
    {\{u_{k_\tau},\cdots,u_{k_\tau+l-1}\}}{\sum_{i=k_\tau}^{k_\tau+l-1} (\bar{g}(\hat{x}_i,\theta_{\tau})^T \tilde{Q}_\tau \bar{g}(\hat x_i,\theta_{\tau}) + u_i^TRu_i) + \bar g(\hat x_l,\theta_{\tau})^T{\tilde{Q}_{l}}\bar g(\hat x_l,\theta_{\tau}) \label{mpckg}}
    {}{}
    \addConstraint{g(\hat x_{i+1},\theta_{\tau}) = A_{\tau}g(\hat x_i,\theta_{\tau}) + B_{\tau}u_i}
    \addConstraint{\hat x_i = C_{\tau}g(\hat x_i,\theta_{\tau})}
    \addConstraint{\hat x_{k_\tau} = x_{k_\tau}}
    \addConstraint{u_i \in \mathcal{U}, \hat x_i \in \mathcal{X}, i = k_\tau,\cdots,k_\tau+\beta_\tau-1}
\end{mini}
where $l \leq \beta_\tau$ is the time horizon, $i$ is the time index along the defined time horizon; $x_i \in \mathbb{R}^{n}$ and $u_i \in \mathbb{R}^{m}$ are the system state and control input at time step $i$, respectively; $\bar{g}(\hat x_i,\theta_{\tau}) = g(\hat x_i,\theta_{\tau})-g(x^*,\theta_{\tau})$ with $x^*=[0,0]'$ denoting the goal state; $\mathcal{X}$ and $\mathcal{U}$ denote the state and control input constraints, respectively; $\tilde{Q}_{\tau} = C_{\tau}^TQC_{\tau} \in \mathbb{R}^{r\times r}$, $Q \in \mathbb{R}^{n\times n}, R \in \mathbb{R}^{m\times m}$ are positive definite matrices.

\begin{figure}[ht]
    \centering
    \includegraphics[width=0.46\textwidth]{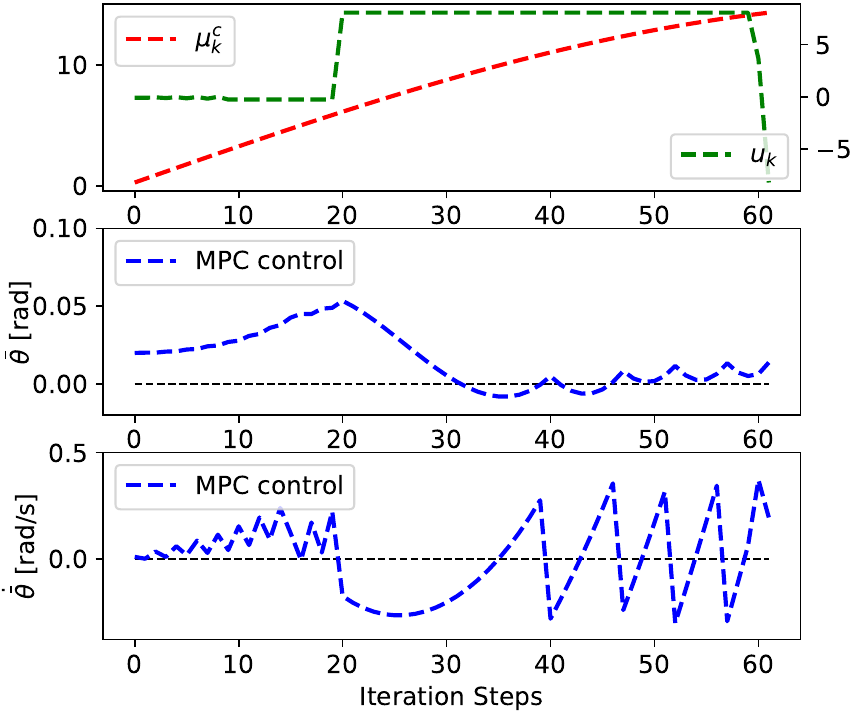}
    \caption{The trajectory generated by the MPC controller using the dynamics learned by DKTV.}\label{fig:traj}
\end{figure}

\textbf{Results Analysis}: Fig. \ref{fig:traj} illustrates the trajectory of the time-varying cartpole system regulated by an MPC controller constructed on the learned deep Koopman model, obtained via the proposed approach. The results demonstrate that the controller successfully maintains the cartpole in the upright position, despite significant variation in the friction coefficient $\mu_k^c$, which increases from $0.0005$ to $14.288$.

\section{Conclusions}\label{conclusion}
This paper presents a data-driven approach to approximate the dynamics of a nonlinear time-varying system (NTVS) by a
linear time-varying system (LTVS), which is resulted from the Koopman operator and deep neural networks. The proposed algorithm is able to approximate an NTVS well by iterative updating its linear approximation, as validated by a simple NTVS system in two dimensions. Moreover, controllers developed based on linear approximation perform well in controlling a quadrotor with complex dynamics. Future work includes employing the proposed method to achieve adaptive autonomy in \cite{WZZS20NIPS} and to serve as robot's dynamics estimation in human-robot teaming \cite{WTZS23TRO,WTDNS23TRO}.
%the enables the DNN observable function to adapt to an approximation of the dynamics of the unknown NTVS when new state-control data pairs become available during the controller execution. The estimation performance of the proposed method is demonstrated by identifying a 2-D NTVS example with slow and rapid variations. The computational efficiency of the proposed method is also validated by estimating more complex quadcopter dynamics. In future work, it is essential to research the optimal control design based on the proposed algorithm and identify the best choice of $\beta_\tau$.
% \bigskip

% \bibliographystyle{plain}
\bibliographystyle{unsrt}
\bibliography{refs.bib, refs1.bib}
% \bibliography{refs1.bib}

%\appendix
\section{Appendix}
% \subsection{Pseudocodes}\label{DKTValg}
\subsection{Pseudocodes}
\DontPrintSemicolon
\begin{algorithm}[ht!]
\SetAlgoLined
\KwIn{$\beta_0$, $\beta_1$, $\cdots$, $\beta_\tau$, $\cdots$}
\KwOut{$ g(\cdot,\theta_{\tau}), A_{\tau}, B_{\tau}, C_{\tau}$, $\tau = 1, 2, \cdots$}
\kwIni{
    \begin{itemize}
        \item Set $k=\beta_0$, $\tau = 0$, and empty matrices $\bX$, $\bar{\bX}$, $\mathbf{U}$ for temporary data storage. 
        \item Initialize $g(\cdot,\theta_{0})$ with random $\theta_0\in\mathbb{R}^q$, and $A_{0}, B_{0}, C_{0}$ by \eqref{lmn}-\eqref{lmn1} based on the observed $\mathcal{B}_0$.
    \end{itemize}}
    \While{$1$}{
    Insert $x_k\in \mathbb{R}^n$, $x_{k+1}\in \mathbb{R}^n$, $u_k\in \mathbb{R}^m$ as new columns into $\mathbf{X, \bar{X}, U}$, respectively.\\
    $k:=k+1$.\\
    % $\bX$.append($x_t$),$\bY$.append($x_{t+1}$),$\mathbf{U}$.append($u_t$)\\
    \uIf{$k=\sum_{i=0}^{\tau+1} \beta_i$}{
    \kwSet{$\tau := \tau + 1$, $\bX_{\tau} := \bX, \bar{\bX}_{\tau} := \bar{\bX}, \mathbf{U}_{\tau} := \mathbf{U}$.}
        \kwUpdate{$A_\tau^\theta$ and $B_\tau^\theta$ are updated by \eqref{tau1}, and $C_\tau^\theta$ is updated by \eqref{tau2}.}
        \kwBuild{Construct $\textbf{L}(\theta_\tau)$ by \eqref{lossf1} and solve \[\theta_{\tau}^* = \arg\min_{\theta_{\tau}\in\mathbb{R}^q}\bL(\theta_{\tau}).\]}
        \kwUpdate{$A_{\tau}\coloneqq A_\tau^{\theta^*}$, $B_{\tau}\coloneqq B_\tau^{\theta^*}$, $C_{\tau}\coloneqq C_\tau^{\theta^*}$.}
        \kwReset{$\bX\coloneqq\bX_\tau[:,\beta_\tau]$, $\bar{\bX}\coloneqq \!\bar{\bX}_\tau[:,\beta_\tau]$, $\mathbf{U}\coloneqq \mathbf{U}_\tau[:,\beta_\tau]$.}
    }}
\caption{Deep Koopman Learning for Time-Varying Systems (DKTV)}
\end{algorithm}

\subsection{Proof of Lemma~\ref{lemma1}}
%\begin{pf}
Similar to the EDMD, we start by solving the least square problem with collected state-control pairs $\mathcal{B}_\tau$ in \eqref{eq_batch}. Define
\begin{equation}
    \begin{aligned}
        \bX_\tau &= [x_{k_\tau}, x_{k_\tau+1},...,x_{k_\tau+\beta_\tau-1}] \in \mathbb{R}^{n\times \beta_\tau},\\
        \bar{\bX}_\tau &= [x_{k_\tau+1}, x_{k_\tau+2},...,x_{k_\tau+\beta_\tau}]\in \mathbb{R}^{n\times \beta_\tau},\\
        \mathbf{U}_\tau &= [u_{k_\tau}, u_{k_\tau+1},...,u_{k_\tau+\beta_\tau-1}] \in \mathbb{R}^{m\times \beta_\tau}. \nonumber
    \end{aligned}
\end{equation}
Given $g(\cdot, \theta_\tau): \mathbb{R}^n \rightarrow \mathbb{R}^r$ with fixed vector $\theta_\tau\in\mathbb{R}^q$, matrices $A_\tau \in \mathbb{R}^{r\times r}, B_\tau \in \mathbb{R}^{r\times m}$ are obtained by solving
\begin{equation}\label{sysid}
    \displaystyle\min_{A_\tau\in\mathbb{R}^{r\times r}, B_\tau\in\mathbb{R}^{r\times m}} \parallel \bar{\bG}_\tau - A_{\tau}\bG_\tau - B_{\tau} \mathbf{U}_\tau \parallel_F^2,
    \nonumber
\end{equation}
of which the unique minimum-norm solution is given by
\begin{equation}\label{prosol1}
    \begin{bmatrix}A_{\tau}^{\theta}, B_{\tau}^{\theta} \end{bmatrix} = \bar{\bG}_\tau \begin{bmatrix} \bG_\tau \\ \mathbf{U}_\tau \end{bmatrix}^\dagger,
\end{equation}
where $\bG_\tau\in \mathbb{R}^{r\times \beta_\tau}, \bar{\bG}_\tau\in \mathbb{R}^{r\times \beta_\tau}$ are defined in \eqref{gstack1}. Set \[ V_{\tau} =  \bar{\bG}_\tau \begin{bmatrix}
{\bG_\tau} \\
{ \mathbf{U}_\tau}\end{bmatrix}^T, G_{\tau}=\begin{bmatrix}
{\bG_\tau}\\  
{\mathbf{U}_\tau}\end{bmatrix} 
\begin{bmatrix}{\bG_\tau} \\{\mathbf{U}_\tau} 
\end{bmatrix}^T, M_{\tau} =[A_\tau^{\theta_\tau},B_\tau^{\theta_\tau}],\]
then \eqref{prosol1} is rewritten as $M_{\tau} = V_{\tau}G_{\tau}^{-1}$.
As we obtain new state-control data pairs denoted by $\bX_{\tau+1}$, $\bar{\bX}_{\tau+1}$ and $\mathbf{U}_{\tau+1}$,
matrices $V_{\tau+1}, G_{\tau+1}$ are updated by
\begin{equation}
    \begin{aligned}
        V_{\tau+1} = \begin{bmatrix} \bar{\bG}_\tau\ \bar{\bG}_{\tau+1} \end{bmatrix} \begin{bmatrix} \bG_\tau &\bG_{\tau+1} \\ \mathbf{U}_\tau &\mathbf{U}_{\tau+1}\end{bmatrix}^T= V_{\tau} + \bar{\bG}_{\tau+1}\begin{bmatrix}{\bG_{\tau+1}} \\{\mathbf{U}_{\tau+1}}\end{bmatrix}^T, \nonumber
    \end{aligned}
\end{equation}
and
\begin{equation}
    \begin{aligned}
        G_{\tau+1} = \begin{bmatrix}
               \bG_\tau&\bG_{\tau+1}\\
               \mathbf{U}_\tau &\mathbf{U}_{\tau+1} 
         \end{bmatrix}\begin{bmatrix}
               \bG_\tau &\bG_{\tau+1}\\
               \mathbf{U}_\tau  &\mathbf{U}_{\tau+1}
         \end{bmatrix}^T = G_{\tau} + \begin{bmatrix}
       {\bG_{\tau+1}}\\  
       {\mathbf{U}_{\tau+1}} 
 \end{bmatrix} \begin{bmatrix}
       {\bG_{\tau+1}}\\  
       {\mathbf{U}_{\tau+1}} 
 \end{bmatrix}^T.
    \end{aligned}
\end{equation}
According to the Sherman–Morrison formula: \[(A+uv^T)^{-1} = A^{-1} - \frac{A^{-1}uv^TA^{-1}}{\bI+v^TA^{-1}u},\]
one has:
\begin{align*}
 M_{\tau+1}&=(V_{\tau} \!+\! \bar{\bG}_{\tau+1}\chi_{\tau+1}^T)({G_{\tau}^{-1} \!\!-\! (G_{\tau}^{-1}\chi_{\tau+1} \chi_{\tau+1}^T G_{\tau}^{-1}})  ({\mathbf{I}_{\beta_{\tau+1}}+\chi_{\tau+1}^T G_{\tau}^{-1}\chi_{\tau+1}})^{-1} )
\\&= M_{\tau} - \lambda_{\tau}M_{\tau}\chi_{\tau+1} \chi_{\tau+1}^T G_{\tau}^{-1} + \lambda_{\tau}\bar{\bG}_{\tau+1}(\lambda_{\tau}^{-1} \!-\! \chi_{\tau+1}^T G_{\tau}^{-1}\chi_{\tau+1})\chi_{\tau+1}^T G_{\tau}^{-1}
\\ &= M_{\tau} - \lambda_{\tau}M_{\tau}\chi_{\tau+1} \chi_{\tau+1}^T G_{\tau}^{-1} + \lambda_{\tau}\bar{\bG}_{\tau+1}\chi_{\tau+1}^T G_{\tau}^{-1} 
\\ &= M_{\tau} + (\bar{\bG}_{\tau+1} - M_{\tau}\chi_{\tau+1}) \lambda_{\tau}\chi_{\tau+1}^T G_{\tau}^{-1},
\end{align*}
where \[ \lambda_{\tau}=(\bI_{\beta_{\tau+1}}+\chi_{\tau+1}^T  G_{\tau}^{-1}\chi_{\tau+1})^{-1}, \chi_{\tau+1} = \begin{bmatrix}
               {\bG_{\tau+1}}\\  
               {\mathbf{U}_{\tau+1}} 
         \end{bmatrix}, \]
and $C_{\tau+1}^\theta \in \mathbb{R}^{n\times r}$ is proved analogically. $\hfill \blacksquare$
%\end{pf}

% \end{pf}

\subsection{Proof of Lemma~\ref{lemma2}}
Consider a DNN with $h\geq 1$ hidden layers and recall that $\bar{\psi}^{h}= [\bar{\psi}_1^{h}, \bar{\psi}_2^{h}, \cdots, \bar{\psi}_{n_h}^{h}]^T\in\mathbb{R}^{n_h}$ denotes its last hidden layer and $\bar \psi^{o}= [\bar{\psi}_1^{o}, \bar{\psi}_2^{o}, \cdots, \bar{\psi}_r^{o}]^T\in\mathbb{R}^r$ denotes its output layer. Then $\bar \psi^{o}$ can be represented by $\bar{\psi}^{h}$ as $$\bar\psi_{i}^o = \bar\psi_{i}^o(\sum_{j = 1}^{n_h}\theta_{i,j}^h \bar\psi_j^{h}),\quad i=1,2,\cdots,r,$$ where $\theta^h \in \mathbb{R}^{r\times n_{h}}$ denotes the weight matrix of the DNN's last hidden layer and $\bar\psi_i^o:\mathbb{R}\rightarrow \mathbb{R}, \bar\psi_j^h:\mathbb{R}\rightarrow \mathbb{R}$ are generally a nonlinear function chosen by the user. 

Define $\phi= [\phi_1, \phi_2, \cdots, \phi_r]^T\in\mathbb{R}^r$ with $$\phi_i = \sum_{j = 1}^{\infty}\theta_{i,j}^h \bar\psi_j^{h}, \quad i = 1, 2, \cdots, r,$$ and $\parallel \phi \parallel = 1$. Let $\Phi = \begin{bmatrix} \phi\\u \end{bmatrix}$ with $u\in \mathbb{R}^m$ denote the control input chosen from the control sequence $\mathbf{u}$. Recall the projection operator from Definition \ref{defi2}. Let $P \coloneqq \begin{bmatrix}
    P_{n_h}^{\mu_1} & \mathbf{0}_{n_h\times m}\\ \mathbf{0}_{m\times n_h} & P_m^{\mu_2}
\end{bmatrix}$ with $\mu_1, \mu_2$ denoting the empirical measurements associated to the $x_1, x_2, \cdots, x_N$ and $u_1, u_2, \cdots, u_N$, respectively. If Assumption \ref{asp4} holds, according to Parseval’s identity (i.e., $\sum_{i=1}^{r}\sum_{j=1}^{\infty} \parallel\theta_{i,j}^h\bar{\psi}_j^{h}\parallel^2=\sum_{i=1}^{r}\sum_{j=1}^{\infty} \parallel\theta_{i,j}^h\parallel^2= 1$), one has
\begin{align*}
    \parallel P\Phi - \Phi \parallel^2 &= \parallel \begin{bmatrix}
        P_{n_h}^{\mu_1}\phi\\
        P_m^{\mu_2} u
    \end{bmatrix}-\begin{bmatrix}
        \phi\\
        u
    \end{bmatrix}\parallel^2 \nonumber
    \\
    &= \parallel \begin{bmatrix} \sum_{j=1}^{n_h} \theta_{1,j}^h  \bar{\psi}_j^{h} - \sum_{j=1}^{\infty} \theta_{1,j}^h  \bar{\psi}_j^{h}\\ \sum_{j=1}^{n_h} \theta_{2,j}^h  \bar{\psi}_j^{h} - \sum_{j=1}^{\infty} \theta_{2,j}^h  \bar{\psi}_j^{h}\\ \vdots \\ \sum_{j=1}^{n_h} \theta_{r,j}^h  \bar{\psi}_j^{h} - \sum_{j=1}^{\infty} \theta_{r,j}^h  \bar{\psi}_j^{h}\\ \mathbf{0}_m
\end{bmatrix}\parallel^2 \nonumber
    \\&= \sum_{i=1}^{r}\sum_{j=n_h +1}^{\infty} \parallel\theta_{i,j}^h \bar{\psi}_j^{h}\parallel^2= \sum_{i=1}^{r}\sum_{j=n_h +1}^{\infty} \parallel\theta_{i,j}^h\parallel^2 \xrightarrow{n_h\rightarrow \infty} 0.\nonumber
\end{align*}
\hfill $\blacksquare$

\subsection{Proof of Lemma~\ref{lemma3}}
Recall the $L_2$ projection and $\mathcal{F}_r$ from Definition \ref{defi2}.
 Before we start the proof, we introduce the following supporting lemma from \cite{korda2018linear}.
\begin{lemma}\label{sulemma}
Let $\hat{\chi}_k = \begin{bmatrix} x_k\\ u_k \end{bmatrix}$ denote the state of the augmented system and $\mu_k$ be the empirical measurements with respect to the points $\hat{\chi}_1, \hat{\chi}_2, \cdots, \hat{\chi}_k$ denoted by $\mu_k = \frac{1}{k}\sum_{i=1}^k \delta_{\hat{\chi}_i}$, where $\delta_{\hat{\chi}_i}$ denotes the Dirac measurement in $\hat{\chi}_i$. Then for any $f\in\mathcal{F}_r$,
\[K_D f = P_r^{\mu_k} \mathcal{K}f = \arg\min_{g\in\mathcal{F}_r}\parallel g - \mathcal{K}f\parallel_{L_2(\mu_k)}.\]
\end{lemma}
\begin{remark}
    Lemma~\ref{sulemma} shows that the approximated Koopman operator $K_D$ is the $L_2$ projection of the Koopman operator $\mathcal{K}$ in $\mathcal{F}_r$ associated with the empirical measure supported in samples $\hat{\chi}_1, \hat{\chi}_2, \cdots, \hat{\chi}_k$.
\end{remark}
Then recall $\Phi, P$ defined in \textbf{Proof of Lemma~\ref{lemma2}} and rewrite $\Phi$ as $\Phi = P\Phi + (I-P)\Phi$.
According to Lemma~\ref{sulemma}, one lets $K_D \coloneqq P\mathcal{K}$ be the approximated sequence of Koopman operators which leads to
\begin{equation}\label{KDP}
    \begin{aligned}
  & \parallel K_D P\Phi - \mathcal{K}\Phi \parallel \\=& \parallel P\mathcal{K}P\Phi - \mathcal{K}\Phi \parallel\\=& \parallel(P-I)\mathcal{K}P\Phi + \mathcal{K}(P-I)\Phi \parallel \\
    \leq  &\parallel (P-I)\mathcal{K}P\Phi \parallel + \parallel \mathcal{K}(P-I)\Phi \parallel \\
  \leq  &\parallel (P-I)\mathcal{K}\Phi \parallel + \parallel (P-I)(\mathcal{K}P - \mathcal{K})\Phi \parallel + \parallel \mathcal{K}\parallel \parallel (P-I)\Phi \parallel.
    \end{aligned}
\end{equation}
According to Lemma~\ref{lemma2}, one has
$\lim_{n_h\rightarrow\infty}\parallel K_D P\Phi - \mathcal{K}\Phi \parallel = 0$. \hfill $\blacksquare$

\subsection{Proof of Theorem~\ref{mainth}}
Abusing some notations, we first show that the estimation error in \eqref{eq_error} is bounded. Recall that given the latest $\mathcal{B}_\tau$ with $\{x_{k-1}, u_{k-1}\}$ its latest data point (i.e., $x_{k-1}\coloneqq x_{k_\tau+\beta_\tau}$, $u_{k-1}\coloneqq u_{k_\tau+\beta_\tau}$), $x_k\in\mathbb{R}^n$ denotes the state of the system of the unknown NTVS evolving from $x_{k-1}\in\mathbb{R}^n$ and $u_{k-1}\in\mathbb{R}^m$. $\hat{x}_k$ in \eqref{xhat} denotes the estimated state introduced by the proposed method. Here, we extend the estimation error in \eqref{eq_error} as 
\begin{equation}\label{error_extend}
 \parallel e_k \parallel = \parallel x_k - \hat{x}_k +  x_{k-1} - x_{k-1}\parallel.
\end{equation}
For any $\bar{x}\in \mathcal{B}_\tau^x, \bar{u}\in \mathcal{B}_\tau^u$ (note that here $x_{k-1}=\bar{x}_{k-1}, u_{k-1}=\bar{u}_{k-1}$ since $\{x_{k-1}, u_{k-1}\}$ is the observed state-input data point), we introduce $\epsilon_k$ as the local estimation error induced by the system approximation in \eqref{pro1} given by
    \[\epsilon_k = g(\bar{x}_{k},\theta_{\tau}) - (A_{\tau}g(\bar{x}_{k-1},\theta_{\tau})+B_\tau \bar{u}_{k-1}).\] 
Similarly, let $\bar{\epsilon}_k$ denote the approximation error induced by the minimization of \eqref{pro11} by
\begin{equation}
    \bar{\epsilon}_k = \bar{x}_k - C_\tau g(\bar{x}_k,\theta_\tau),\nonumber
\end{equation}
which leads to
\begin{equation}\label{eq_xbarref}
    \begin{aligned}
        \bar{x}_{k-1} =  C_\tau(A_\tau g(\bar{x}_{k-2}, \theta_\tau) + B_\tau \bar{u}_{k-2}) + C_\tau \epsilon_{k-1} + \bar{\epsilon}_{k-1}.
    \end{aligned}
\end{equation}
By substituting \eqref{xhat} and \eqref{eq_xbarref} into \eqref{error_extend}, we have the following.
\begin{equation}\label{p1p}
\begin{aligned}
    \parallel e_k \parallel =& \parallel C_\tau A_\tau (g(\bar{x}_{k-2}, \theta_\tau) -g(\bar{x}_{k-1}, \theta_\tau)) + C_\tau B_\tau (\bar{u}_{k-2} - \bar{u}_{k-1})  \\&+ C_{\tau}\epsilon_{k-1} + x_k - \bar{x}_{k-1} + \bar{\epsilon}_{k-1} \parallel,\\
     \overset{(i)}{\le}& \parallel C_\tau A_\tau (g(\bar{x}_{k-2}, \theta_\tau) -g(\bar{x}_{k-1}, \theta_\tau))\parallel + \parallel C_\tau B_\tau(\bar{u}_{k-2} - \bar{u}_{k-1})\parallel  \\&+ \parallel C_{\tau}\epsilon_{k-1}\parallel + \parallel x_k - \bar{x}_{k-1}\parallel + \parallel \bar{\epsilon}_{k-1} \parallel,\\
    \overset{(ii)}{\le} &\parallel C_\tau A_\tau \parallel \parallel g(\bar{x}_{k-2}, \theta_\tau) -g(\bar{x}_{k-1}, \theta_\tau)\parallel +  \parallel C_\tau B_\tau \parallel \parallel \bar{u}_{k-2} - \bar{u}_{k-1}\parallel  \\&+ \parallel C_{\tau}\parallel \parallel\epsilon_{k-1}\parallel + \parallel x_k - \bar{x}_{k-1}\parallel + \parallel \bar{\epsilon}_{k-1} \parallel,\\
\end{aligned}
\end{equation}
where $(i)$ follows the triangle inequality and $(ii)$ is derived by subordinance and submultiplicativity.

For the first two terms of (\ref{p1p}), if Assumptions \ref{asp2}-\ref{asp3} hold, one has
\[\parallel C_\tau A_\tau \parallel \parallel g(\bar{x}_{k-2}, \theta_\tau) - g(\bar{x}_{k-1}, \theta_\tau)\parallel \leq \parallel C_\tau A_\tau \parallel \mu_g \mu_x\] and \[\parallel C_\tau B_\tau \parallel \parallel \bar{u}_{k-2} - \bar{u}_{k-1}\parallel \leq \parallel C_\tau B_\tau \parallel \mu_u .\]
For writing convenience, one denotes 
\begin{equation}\label{firstbd}
    L_a \coloneqq \parallel C_\tau A_\tau \parallel \mu_g \mu_x + \parallel C_\tau B_\tau \parallel \mu_u.
\end{equation}
For the third term of (\ref{p1p}), one can derive the error bound of $\parallel \epsilon_{k-1} \parallel$ by backpropagating it from $k$ to $k_0=0$. Define the global approximation error in $\mathcal{B}_\tau$ as
\begin{equation}\label{cet}
        \mathcal{E}_k =  g(\bar{x}_{k},\theta_{\tau}) - \prod_{i=0}^{k}A_i g(\bar{x}_0,\theta_0)-\sum_{j=0}^{k-1} (\prod_{l=0}^{k-j-1}A_l)B_j u_j .
\end{equation}
The related proof is inspired by the global accumulation rule of $\epsilon_k$ of time-invariant systems proposed in \cite{mamakoukas2020learning} given by
\begin{equation}\label{ideae}
        E_k =  \sum_{i=0}^{k-1}A^i\epsilon_{k-i}.
\end{equation}
To achieve the error bound $\mathcal{E}_k$ in \eqref{cet}, it is necessary to replace $A$ in \eqref{ideae} with $A_\tau$ (note that the batch index $\tau$ is slower than the data point index $k$). Then $\mathcal{E}_k$ is derived by induction, that is,
\begin{itemize}
    \item when $\tau = 1,  k\in [k_1, k_1+\beta_1]$,
    \item[] \begin{equation}\label{cet1}
    \begin{aligned}
    \mathcal{E}_{k} = \sum_{i=0}^{k-k_1-1} A_{1}^{i}\epsilon_{k-i} + A_{1}^{k-k_1-1}\sum_{j=1}^{\beta_{0}}A_{0}^j\epsilon_{k_1+1-j}, \end{aligned}
    \end{equation}
    \item when $\tau>1, k\in [k_\tau, k_\tau+\beta_\tau]$,
    \item[] \begin{equation}\label{cet3}
    \begin{aligned}
    \mathcal{E}_{k} = \sum_{i=0}^{k-k_\tau-1} A_{\tau}^{i}\epsilon_{k-i} + A_{\tau}^{k-k_\tau-1} \Bigg(\sum_{j=1}^{\beta_{\tau-1}}A_{\tau-1}^j
    \epsilon_{k_\tau+1-j}+\\ \Big(\sum_{l=1}^{\tau-1}\sum_{m=1}^{\beta_{l-1}}(\prod_{n=\tau-1}^{l} A_{n}^{\beta_{n}}) A_{l-1}^{m}\Big)
     \epsilon_{k_l+1-m}\Bigg),
    \end{aligned}
    \end{equation}
    \item when $\tau+1, k\in [k_{\tau+1}, k_{\tau+1}+\beta_{\tau+1}]$,
    \item[] \begin{equation}\label{cet4}
    \begin{aligned}
    \mathcal{E}_{k} = \sum_{i=0}^{k-k_{\tau+1}-1} A_{\tau+1}^{i}\epsilon_{k-i} + A_{\tau+1}^{k-k_{\tau+1}-1} \Bigg(\sum_{j=1}^{\beta_{\tau}} A_{\tau}^j\epsilon_{k_{\tau+1}+1-j}\\ +
    \Big(\sum_{l=1}^{\tau}\sum_{m=1}^{\beta_{l-1}}(\prod_{n=\tau}^{l} A_{n}^{\beta_{n}}) A_{l-1}^{m}\Big)\epsilon_{k_l+1-m}\Bigg).
    \end{aligned}
    \end{equation}
\end{itemize}
According to \eqref{cet1}-\eqref{cet4}, the third term of (\ref{p1p}) is bounded by
\begin{equation}
    \begin{aligned}
    \parallel C_{\tau}\epsilon_{k-1} \parallel \leq
    \parallel C_{\tau}\Bigg(\sum_{i=0}^{k-k_\tau-2} A_{\tau}^{i} + A_{\tau}^{k-k_\tau-2} \bigg(\sum_{j=1}^{\beta_{\tau-1}}A_{\tau-1}^j \\+ \sum_{l=1}^{\tau-1}\sum_{m=1}^{\beta_{l-1}}(\prod_{n=\tau-1}^{l} A_{n}^{\beta_{n}}) A_{l-1}^{m}\bigg)\Bigg)\parallel L_1 \nonumber.
    \end{aligned}
\end{equation}
where 
\begin{equation}\label{L111}
    \begin{aligned}
    L_1 = \max_{\overset{\bar{x}_{s}\in\mathcal{B}_\tau^x,} {\bar{u}_s\in\mathcal{B}_\tau^u}} \parallel g(\bar{x}_{s+1},\theta_{\tau}) - A_{\tau}g(\bar{x}_s,\theta_{\tau}) - B_\tau \bar{u}_s \parallel.
    \end{aligned}
\end{equation}

For the last two terms of (\ref{p1p}), if Assumption \ref{asp2} holds, one has
\begin{equation}
    \begin{aligned}
  \parallel x_k - \bar{x}_{k-1} \parallel \leq\mu_x, \nonumber
    \end{aligned}
\end{equation}
and since $\bar{\epsilon}_{k-1}$ is defined on $\mathcal{B}_\tau^x$, one can compute its upper bound by
\begin{equation}\label{L222}
    \parallel \bar{\epsilon}_{k-1} \parallel \leq \max_{\bar{x}\in\mathcal{B}_\tau^x}\parallel \bar{x} - C_\tau g(\bar{x},\theta_\tau) \parallel \eqqcolon L_2.
\end{equation}
To sum up, recall $L_1$ in \eqref{L111} and $L_2$ in \eqref{L222}, let
\begin{equation}\label{Lbbound}
    \begin{aligned}
    L_b \coloneqq 
   L_1 \parallel C_{\tau}\Bigg(\sum_{i=0}^{k-k_\tau-2} A_{\tau}^{i} + A_{\tau}^{k-k_\tau-2} \bigg(\sum_{j=1}^{\beta_{\tau-1}}A_{\tau-1}^j + \sum_{l=1}^{\tau-1}\sum_{m=1}^{\beta_{l-1}}(\prod_{n=\tau-1}^{l} A_{n}^{\beta_{n}}) A_{l-1}^{m}\bigg)\Bigg)\parallel 
    \end{aligned}
\end{equation}
and 
\begin{equation}\label{Lcbound}
    \begin{aligned}
  L_c \coloneqq \mu_x + L_2, 
    \end{aligned}
\end{equation}
the estimation error $e_k$ in \eqref{error_extend} is upper bounded by 
\begin{equation}\label{eq_errorbound}
    \parallel e_k \parallel \leq L_a + L_b + L_c.
\end{equation}
We further reduce the error bound derived in \eqref{eq_errorbound} based on Lemmas \ref{lemma2}-\ref{lemma3}. Recall that $K_D \coloneqq P \mathcal{K}$ in \eqref{KDP} and let $$\Phi_D^\tau(\bar{x}_k, \bar{u}_k) \coloneqq [g(\bar{x}_k,\theta_{\tau})^T, \bar{u}_k^T]^T.$$
According to Lemma~\ref{lemma2} ($\lim_{n_h\rightarrow\infty}\Phi_D^\tau = \Phi^\tau$) and Lemma~\ref{lemma3} ($\lim_{n_h\rightarrow\infty}K_D^\tau = \mathcal{K}^\tau$), by following the Definition of Koopman operator in \ref{defi1}, one has
\begin{equation}\label{eq_thmmm}
    \begin{aligned}
        \lim_{n_h\rightarrow \infty} K_D^\tau \Phi_D^\tau(\bar{x}_s, \bar{u}_s)\! =\! [\mathcal{K}^\tau \Phi^\tau](\bar{x}_s, \bar{u}_s) \! =\! \Phi^\tau(\bar{x}_{s+1}, \bar{u}_{s+1}).
    \end{aligned}
\end{equation}
Since we are interested in predicting the future values of the system state, we can only keep the first $r$ rows of $\Phi_D^\tau(\bar{x}_{s+1}, \bar{u}_{s+1})$ and $K_D^\tau \Phi_D^\tau(\bar{x}_s, \bar{u}_s)$. Therefore, we define $\bar{\mathcal{K}}^\tau$ and $\bar{K}_D^\tau$ as the first $r$ rows of $\mathcal{K}^\tau$ and $K_D^\tau$, respectively, and decompose the matrix $\bar{K}_D^\tau$ as $\bar{K}_D^\tau \eqqcolon [A_\tau, B_\tau]$. Then according to \eqref{eq_thmmm}, one has
$$ \lim_{n_h\rightarrow \infty} L_1 = \max_{\overset{\bar{x}_{s}\in\mathcal{B}_\tau^x,} {\bar{u}_s\in\mathcal{B}_\tau^u}}\parallel  g(\bar{x}_{s+1},\theta_{\tau}) - [\bar{\mathcal{K}}^\tau \Phi^\tau](\bar{x}_s, \bar{u}_s) \parallel = 0. $$ 

Finally, recall $L_a$ in \eqref{firstbd} and $L_c$ in \eqref{Lcbound}, one has
\begin{equation}
\begin{aligned}
    \lim_{n_h\rightarrow\infty} \text{sup}\parallel e_k\parallel =  (\parallel C_\tau A_\tau \parallel \mu_g + 1) \mu_x + \parallel C_\tau B_\tau \parallel  \mu_u + \max_{\bar{x}\in\mathcal{B}_\tau^x}\parallel \bar{x} - C_\tau g(\bar{x},\theta_\tau) \parallel, \nonumber 
\end{aligned}
\end{equation}
where $\{g(\cdot,\theta_\tau), A_\tau, B_\tau, C_\tau\}$ denotes the achieved the DKR using the proposed DKTV method.
\hfill $\blacksquare$

\subsection{Proof of Corollary~\ref{cor1}}
Recall $L_b$ in \eqref{Lbbound}. If Assumption \ref{asp5} holds, by following the properties of triangle inequality and submultiplicativity, one has $\parallel A_{\tau}^{\beta_\tau} \parallel  < 1$ and $ \parallel\sum_{i=0}^{\beta_\tau} A_{\tau}^i \parallel  < \beta_\tau$, which leads to
\begin{equation}
    \begin{aligned}
    L_b <
    \parallel C_\tau \parallel L_1 (\beta_\tau \!+\! \beta_{\tau-1} +\! \parallel\! \sum_{l=1}^{\tau-1}\sum_{m=1}^{\beta_{l-1}}(\!\prod_{n=\tau-1}^{l}\! A_{n}^{\beta_{n}}) A_{l-1}^{m})\parallel).\nonumber
    \end{aligned}
\end{equation}
Following the submultiplicativity, one has $$\lim_{\tau\rightarrow\infty}\parallel \sum_{l=1}^{\tau-1}\sum_{m=1}^{\beta_{l-1}}(\prod_{n=\tau-1}^{l} A_{n}^{\beta_{n}}) A_{l-1}^{m})\parallel = \mu_c$$ with $\mu_c$ a positive constant, which leads to 
\begin{equation}\label{eq_corp1}
    \lim_{\tau\rightarrow\infty} \text{sup} L_b = \parallel C_\tau \parallel L_1 (\beta_\tau + \beta_{\tau-1}+\mu_c ).
\end{equation}
By substituting \eqref{eq_corp1} into \eqref{eq_errorbound}, one can notice that $\lim_{k\rightarrow\infty} \sup  \parallel e_k\parallel$ is determined by the minimization performance of \eqref{lossf1} and $\mu_x, \mu_u$. $\hfill \blacksquare$
% \end{pf}








% \end{thebibliography}
\end{document}